\documentclass[iop,onecolumn,numberedappendix]{emulateapj}
\usepackage{graphicx,amsmath,bbold,dsfont,bbm,natbib}
\slugcomment{Submitted to ApJ}

\shorttitle{First order TTVs}
\shortauthors{Agol \& Deck}

\begin{document}

\title{Transit timing to first order in eccentricity}
\author{Eric Agol}

\email{agol@uw.edu}

\affil{Astronomy Department, University of Washington, Seattle, WA 98195; Kavli Institute for Theoretical Physics, University of California, Santa Barbara, CA 93106, USA; NASA Astrobiology Institute’s Virtual Planetary Laboratory, Seattle, WA 98195, USA}

\and

\author{Katherine Deck}
\affil{Division of Geological and Planetary Sciences, Caltech, 1200 E. California Blvd., Pasadena, CA 91125 USA}

\date{August 2015}

\begin{abstract}
Characterization of transiting planets with transit timing
variations (TTVs) requires understanding how to translate the observed
TTVs into masses and orbital elements of the planets.  This
can be challenging in multi-planet transiting systems, but fortunately these
systems tend to be nearly plane-parallel and low eccentricity.
Here we present a novel derivation of analytic formulae for TTVs 
that are accurate to first order in the planet-star mass ratios and in the
orbital eccentricities.  These formulae are accurate in proximity to 
first order resonances, as well
as away from resonance, and compare well with
more computationally expensive N-body integrations in the low
eccentricity, low mass-ratio regime when applied to simulated and
to actual multi-transiting Kepler planet systems.  We make code available for
implementing these formulae.
\end{abstract}


\section{Introduction}

No planet orbits on a precisely Keplerian orbit:  post-Newtonian
corrections, stellar oblateness, and, most importantly, planetary perturbations cause deviations
from a periodic ephemeris for transiting exoplanets \citep{Miralda2002,
Schneider2003,Schneider2004,Holman2005,Agol2005,Heyl2007,NesvornyMorbidelli2008,Fabrycky2010}.
Transit-timing variations (TTVs) have been used to confirm that
transit signals are in fact due to planets \citep{Kepler9,Ford2012,Steffen2012,confirm1,confirm2,confirm3,Xie2013,Xie2014a},
to detect and characterize non-transiting planets \citep{Ballard,KOI872,Nesvorn2013}, and to
make precise measurements of the masses and dynamical states of
multi-transiting exoplanet systems \citep[e.g.][]{CarterAgol2012}.

For the latter two applications, fast computation of TTVs is required
for rapid searching through parameter space for perturbing companions,
and for rapid computation of the posterior distributions of the masses
and orbital elements of transiting planet systems.   Numerical
computation of TTVs can be sped up through symplectic integration, 
through a more efficient numerical solution of Kepler's equation, and 
through transit time interpolation \citep{TTVFast}; however, this approach 
can still be too computationally intensive for high multiplicity systems,
and does not pinpoint the physical origin of constraints upon
planetary system properties. Analytic formulae based on perturbation theory can greatly speed 
computation, but the perturbation theory to high order in eccentricity
and inclination becomes complicated quickly, and numerical codes that implement the 
analytic formulae have not been released or widely used 
\citep{NesvornyMorbidelli2008,Nesvorn2009,Nesvorn2010}. 
Much can be accomplished with first order (in eccentricity) perturbation
theory because orbital eccentricities of many planets exhibiting TTVs are
small.  TTVs are most easily observed for pairs of planets near a mean
motion resonance; thanks to the low eccentricity of the systems in
consideration \citep{Fabrycky2014,Hadden2014,Limbach2014,VanEylen2015}, the first order resonances are most represented among TTV
pairs, and it is for first order resonances that a first order theory is
adequate.  TTVs caused by these first order resonant interactions are
primarily sinusoidal and are subject to a degeneracy between mass and eccentricity \citep{Boue2012},
caused by mixing of two frequencies of perturbation which are aliased
at the frequency of the transiting planet, as explained in an elegant 
analysis by \citet[][hereafter LXW12]{Lithwick2012}.  To break this degeneracy requires 
the measurement of additional modes, such as the short-timescale TTVs known as 
`chopping' variations \citep{Kepler9,Deck2015}, or statistical analysis
of many systems \citep{LW1,Hadden2014,Xie2014b}.  In addition, the
LXW12 analysis is only approximate, and breaks down for pairs further
from resonance \citep{Deck2015}.  These issues motivate the current
paper in which we derive an explicit formula for TTVs accurate
to first order in eccentricity and planet-star mass ratio, valid
for (nearly) plane-parallel transiting planets (although \citealt{Nesvorny2014}
showed that mutual inclinations of planets can be large and still be
well described by coplanar TTVs).

We expect that these results will be useful a) for determining
how different frequencies within the TTV signal constrain the planetary
masses and orbital elements \citep{Deck2015}; b) for analyzing
systems with a large number of interacting transiting planets
by making linear additions of the analytic formula for pairs of
planets \citep{Kepler11,Jontoff2014}; and c) for rapid search through
parameter space of perturbing planets.

We first summarize the TTV solution to first order in eccentricity and mass in \S \ref{ttv_result} (the full derivation is given in appendix \ref{ttv_calculation}).
We then compare these with prior results, both analytic and numeric
(\S \ref{ttv_comparison}), including comparison of analytic and numeric 
analyses of specific systems.  We discuss the numerical implementation
and speed in \S \ref{numerics}.  We end with a discussion of the
possible applications and future directions (\S \ref{conclusions}).


\section{First-order solution:} \label{ttv_result}

Here we give a complete summary of the assumptions and variables used, and the 
solution to the first-order equations for readers that wish to simply use the 
results of this computation.  The details of the derivation are given in
Appendix \ref{ttv_calculation}.
Since multi-planet transiting systems typically have nearly edge-on orbits 
and hence small mutual inclinations, it is usually
sufficient in analytic approximations to treat the problem in the plane-parallel
approximation.  This leaves four orbital elements for each planet (semi-major
axis, $a_i$, mean longitude,
$\lambda_i$, eccentricity, $e_i$, and longitude of periastron, $\varpi_i$), plus the mass
ratio of each planet to the star, $m_1/m_\star,m_2/m_\star$, where $m_1$ is the mass of
the inner planet, $m_2$ is the mass of the outer planet, and $m_\star$ is the mass 
of the star.  For nearly circular 
planetary orbits, there are two small dimensionless parameters in the problem: $\mu_i=m_i/m_\star$ and $e_i$.
The usual procedure for computing transit timing variations is to: 1) to write
down a Hamiltonian (or disturbing function) for perturbations due to another
planet; 2) expand the Hamiltonian as a function of
the orbital elements to the order in eccentricity desired plus one (e.g.\ if 
a transit timing solution is needed to first order in eccentricity, then
the Hamiltonian must be expanded to second order in eccentricity), including
the linear combinations of mean-longitudes leading to the important resonant
terms necessary for sufficient accuracy; 3) compute
the variation in the orbital elements using Hamilton's equations, which
are four first-order partial differential equations for each planet, and
involves differentiating the Hamiltonian with respect to the orbital elements
(which can be a rather complex operation);
4) integrate the resulting equations as a function of time; 5) compute
the true longitudes, $\theta_i = \theta_{i,K} + \delta \theta_i$,
as a function of time, where $\theta_{i,K}$ is the unperturbed
Keplerian orbit, and $\delta \theta_i$ is the perturbation of
the $i$th planet caused by its planet companion(s); 6) compute the transit timing variations:
\begin{equation}\label{ttv}
\delta t_i =-\dot\theta_{i,K}^{-1}  \delta \theta_i.
\end{equation}  This is the approach
taken by \citet{Agol2005}, \citet{Nesvorny2014}, and LXW12. 
A different approach employing Hamiltonian perturbation theory 
\citep{NesvornyMorbidelli2008} was used in \citet{Deck2015}. This involves 
determining the canonical transformation between the full canonical orbital 
element set and the average set; the TTVs, which are deviations from an 
average ``Keplerian" orbit, can be derived from this transformation.

The standard procedure outlined in detail above (based on Hamilton's equations) has the advantages 
of requiring only first-order differential
equations for the computation, and the advantage of using standard methods
in celestial mechanics for the computation.  However, there are two
possible drawbacks: 1) the expansion of the Hamiltonian in orbital elements 
can be rather complex; 2) the main quantity of interest for transit-timing 
variations is $\delta \theta_i$, rather than the perturbed orbital elements. The derivation based on canonical transformations \citep{NesvornyMorbidelli2008}, though elegant, has the disadvantage of requiring the extra machinery  and knowledge of Hamiltonian perturbation theory.

In our new derivation we forgo computing
the orbital elements, and simply treat the problem in polar coordinates
$(r_i,\theta_i)$.  We then use Newton's equations in terms
of a disturbing function which can be expressed as a function of polar
coordinates, with the added advantage that Newton's equations make
clearer which forces are causing the perturbations. 
This approach has some possible advantages: 1) only two
differential equations are necessary (albeit second-order rather
than first-order); 2) the derivatives of the disturbing function with
respect to the polar coordinates are easy to compute; 3) the
perturbed polar coordinates directly yield the transit timing variations;
4) the resulting expression is more compact than in the Hamiltonian
formulation.
The second-order differential equation may seem like a drawback, but
it can be solved using complex notation (as in LXW12) and by
expanding the derivatives of the disturbing function in terms of orbital
elements, which yields harmonic functions which are easy
to integrate.  The final answer is expressed as a sum over
harmonics of the perturbing planet's orbital frequency \citep{Deck2015}.
Each coefficient for each planet in the harmonic series solution
can be solved for by inverting three two-by-two matrices, which have
a standard format, resulting directly in the transit timing variations 
at a particular frequency.

The unperturbed orbital frequencies, $n_i=2\pi/P_i$, are defined by $n_i^2=Gm_\star/a_i^3$.
As usual, $\alpha = a_1/a_2 \approx (P_1/P_2)^{2/3}$.
We define
$\tilde{A}_{jmn}=a_1^ma_2^{n+1}\frac{\partial^{m+n}}{\partial a_1^m \partial a_2^n}\left(a_2^{-1}b_{1/2}^{(j)}(\alpha)\right)$
where $b_{1/2}^{(j)}(\alpha)$ is the Laplace coefficient,
\begin{equation}
b_{1/2}^{(j)}(\alpha) = \frac{1}{\pi}\int_0^{2\pi} d\theta \frac{\cos{(j \theta)}}{(1+\alpha^2-2\alpha
\cos{\theta})^{1/2}}.
\end{equation}
The difference in mean longitude of the planets is $\psi = \lambda_1-\lambda_2$.
Auxiliary dimensionless quantities are:
\begin{align}
\beta_j &= j (n_1-n_2)/n_1 = j(1-\alpha^{3/2}),\cr
\kappa_j &= j(n_1-n_2)/n_2 = j(\alpha^{-3/2}-1).
\end{align}
The functions $\tilde{A}_{jmn}$ we use below are given by:
\begin{eqnarray}
\tilde{A}_{j00}&=&b_{1/2}^{(j)}(\alpha),\cr
\tilde{A}_{j10}&=&\alpha \partial b_{1/2}^{(j)}/\partial \alpha,\cr
\tilde{A}_{j20}&=&\alpha^2 \partial^2 b_{1/2}^{(j)}/\partial \alpha^2,\cr
\tilde{A}_{j01}&=&-(\tilde{A}_{j10}+\tilde{A}_{j00}) = -(\alpha\partial b_{1/2}^{(j)}/\partial \alpha + b_{1/2}^{(j)}),\cr
\tilde{A}_{j02}&=&2\tilde{A}_{j00}+4\tilde{A}_{j10}+\tilde{A}_{j20} = 2 b_{1/2}^{(j)} + 4\alpha\partial b_{1/2}^{(j)}/\partial \alpha + \alpha^2\partial^2 b_{1/2}^{(j)}/\partial \alpha^2,\cr
\tilde{A}_{j11} &=& -(2\tilde{A}_{j10} +\tilde{A}_{j20}) = -2 \alpha \partial b_{1/2}^{(j)}/\partial \alpha -\alpha^2 \partial^2 b_{1/2}^{(j)}/\partial \alpha^2.
\end{eqnarray}

To use this solution in computing TTVs, the longitudes need to be computed
from the observed transit times; the mean ephemeris, $(t_{0,i},P_i)$, may be used
in computing the (unperturbed) orbital ephemeris.
Now, the mean longitudes are given to first order in eccentricity by
\begin{equation}\label{longitude}
\lambda_i = 2\pi \left(\frac{t-t_{0,i}}{P_i}\right) + 2 e_i \sin{\varpi_i},
\end{equation}
if we assume that the orbital reference is along the line of
sight, and thus $\lambda_i \approx 0$ at the times of transit.

The solutions for the inner planet ($i=1$) and outer planet ($i=2$) are given by:
\begin{eqnarray}\label{timing_solution}
\delta t_1
&=& \frac{P_1}{2\pi} \mu_2 \sum_{j\ge 1} \Big[f_{1,j}^{(0)} \sin{(j\psi)}
+ f_{1,j}^{(-1)}e_1 \sin{\left[j\psi-(\lambda_1-\varpi_1)\right]}
+f_{1,j}^{(+1)}e_1 \sin{\left[j\psi+(\lambda_1-\varpi_1)\right]}\cr
&+&f_{1,j-1}^{(-2)}e_2\sin{\left[j\psi-(\lambda_1-\varpi_2)\right]}
+f_{1,j+1}^{(+2)}e_2 \sin{\left[j\psi+(\lambda_1-\varpi_2)\right]}\Big],\cr
\delta t_2 &=& \frac{P_2}{2\pi} \mu_1 \sum_{j\ge 1} \Big[f_{2,j}^{(0)} \sin{(j\psi)}
+f_{2,j}^{(-2)} e_2 \sin{\left[j\psi-(\lambda_2-\varpi_2)\right]}
+f_{2,j}^{(+2)} e_2 \sin{\left[j\psi+(\lambda_2-\varpi_2)\right]}\cr
&+&f_{2,j+1}^{(-1)} e_1 \sin{\left[j\psi-(\lambda_2-\varpi_1)\right]}
+f_{2,j-1}^{(+1)} e_1 \sin{\left[j\psi+(\lambda_2-\varpi_1)\right]}\Big],
\end{eqnarray}
where the functions $f_{i,j}^{(\pm k)}$ are given by:
\begin{eqnarray}
f_{i,j}^{(\pm k)}(\alpha) &=& u(\gamma,c_1,c_2) + \delta_{ik} v_\pm (\zeta,d_1,d_2),\cr
u(\gamma,c_1,c_2) &=& \frac{\left(3+\gamma^2\right)c_1 + 2\gamma c_2}{\gamma^2\left(1-\gamma^2\right)}\cr
v_\pm (\zeta,d_1,d_2) &=& \frac{\left(\pm\left(1-\zeta^2\right)+6\zeta\right)d_1 + \left(2+\zeta^2\right) d_2}{\zeta(1-\zeta^2)(\zeta\pm 1)(\zeta\pm 2)}.
\end{eqnarray}
where $\gamma$, $c_1$, and $c_2$
and $\zeta$, $d_1$, and $d_2$ are given in Table \ref{tab01}, and $\delta_{ik}$ is the
Kronecker delta function.
Note that the top signs in $\pm,\mp$ correspond to $+k$ values, while the bottom correspond to $-k$.
The functions $f_{i,j}^{(\pm k)}$ are solely a function of $j$, $\pm k$, and $\alpha$.

\begin{table}
\centering
\caption{Coefficients for $u(\gamma,c_1,c_2)$ and $v_\pm(\zeta,d_1,d_2)$ in first-order TTV solution (the $v_\pm$ coefficients
correspond to the $k$ values in brackets [..]).}
\begin{tabular}{cclll}\label{tab01}
$i$ & $\pm k$ & $\gamma [\zeta]$     & $c_1 [d_1] $ & $c_2 [d_2]$ \cr
\hline
1   & $0[\pm 1]$       & $\beta_j$      & $\alpha j\left(\tilde{A}_{j00}-\alpha\delta_{j1}\right)$ & $\alpha\left(\tilde{A}_{j10}-\alpha\delta_{j1}\right)$ \cr
1   & $\pm$1  & $\beta_j\pm 1$ & $\alpha j\left(\pm j\tilde{A}_{j00}-\frac{1}{2}\tilde{A}_{j10}+\frac{1}{2}(1\mp 2)\alpha\delta_{j1}\right)$ & $\alpha\left(\pm j\tilde{A}_{j10}-\frac{1}{2}\tilde{A}_{j20}\mp \alpha\delta_{j1}\right)$ \cr
1   & $\pm$2  & $\beta_j\pm\alpha^{3/2}$   & $\alpha j\left(\mp j\tilde{A}_{j00}-\frac{1}{2}\tilde{A}_{j01}-(1\mp 1)\alpha\delta_{j1}\right)$ & $\alpha\left(\mp j\tilde{A}_{j10}-\frac{1}{2}\tilde{A}_{j11}-(1\mp 1)\alpha\delta_{j1}\right)$\cr
\hline
2   & $0[\pm 2]$       & $\kappa_j$     & $-j\left(\tilde{A}_{j00}-\alpha^{-2}\delta_{j1}\right)$ & $\tilde{A}_{j01}-\alpha^{-2}\delta_{j1}$\cr
2   & $\pm 1$ & $\kappa_j\pm\alpha^{-3/2}$  & $-j\left(\pm j\tilde{A}_{j00}-\frac{1}{2}\tilde{A}_{j10}-(1\pm 1)\alpha^{-2}\delta_{j1}\right)$ &
$\pm j\tilde{A}_{j01} -\frac{1}{2}\tilde{A}_{j11}-(1\pm 1)\alpha^{-2}\delta_{j1}$\cr
2   & $\pm 2$ & $\kappa_j\pm 1$& $-j\left(\mp j\tilde{A}_{j00}-\frac{1}{2}\tilde{A}_{j01}+\frac{1}{2}(1\pm 2)\alpha^{-2}\delta_{j1}\right)$ &
$\mp j\tilde{A}_{j01}-\frac{1}{2}\tilde{A}_{j02}\pm\alpha^{-2}\delta_{j1}$\cr
\hline
\end{tabular}
\end{table}

In practice the sum over $j$ from 1 to $\infty$ must be truncated at
a finite value of $j_{max}$.  Typically $j_{max}$ does not need to
be chosen to be too large since the Laplace coefficients decline
in amplitude with $j$ \citep{Deck2015}.  We recommend choosing a
$j_{max}$ large enough such that the resulting computation is
converged.

As an example of using Table \ref{tab01}, the coefficient
$f_{1,j}^{(-1)}$ has $i=k=1$, $\gamma=\beta_j-1$, $c_1 = 
\alpha j\left(-j\tilde{A}_{j00}-\frac{1}{2}\tilde{A}_{j10}+\frac{3}{2}\alpha\delta_{j1}\right)$
and $c_2 = \alpha\left(-j\tilde{A}_{j10}-\frac{1}{2}\tilde{A}_{j20}+ \alpha\delta_{j1}\right)$,
$\zeta = \beta_j$, $d_1 = \alpha j\left(\tilde{A}_{j00}-\alpha\delta_{j1}\right)$,
and $d_2 = \alpha\left(\tilde{A}_{j10}-\alpha\delta_{j1}\right)$.
Then, the coefficient is given by:
\begin{equation}
f_{1,j}^{(-1)}(\alpha) = u(\beta_j-1,c_1,c_2) + v_-(\beta_j,d_1,d_2).
\end{equation}

\section{Comparison with other formulae}\label{ttv_comparison}

The zeroth-order solution (in the limit $e_1=e_2=0$) compares exactly 
with the results given in \cite{Agol2005}, \citet{Nesvorny2014}, 
and \citet{Deck2015} which were derived with Hamilton's equations 
and the approach based on canonical transformations;  this is reassuring given 
the very different approach used in this derivation.  We have also 
rederived the first-order eccentricity equations using the approach based 
on canonical perturbation theory employed in \cite{Deck2015}, and found exact agreement
with the results presented here to first order in eccentricity.

\begin{figure}
\centering
\includegraphics[width=0.8\hsize]{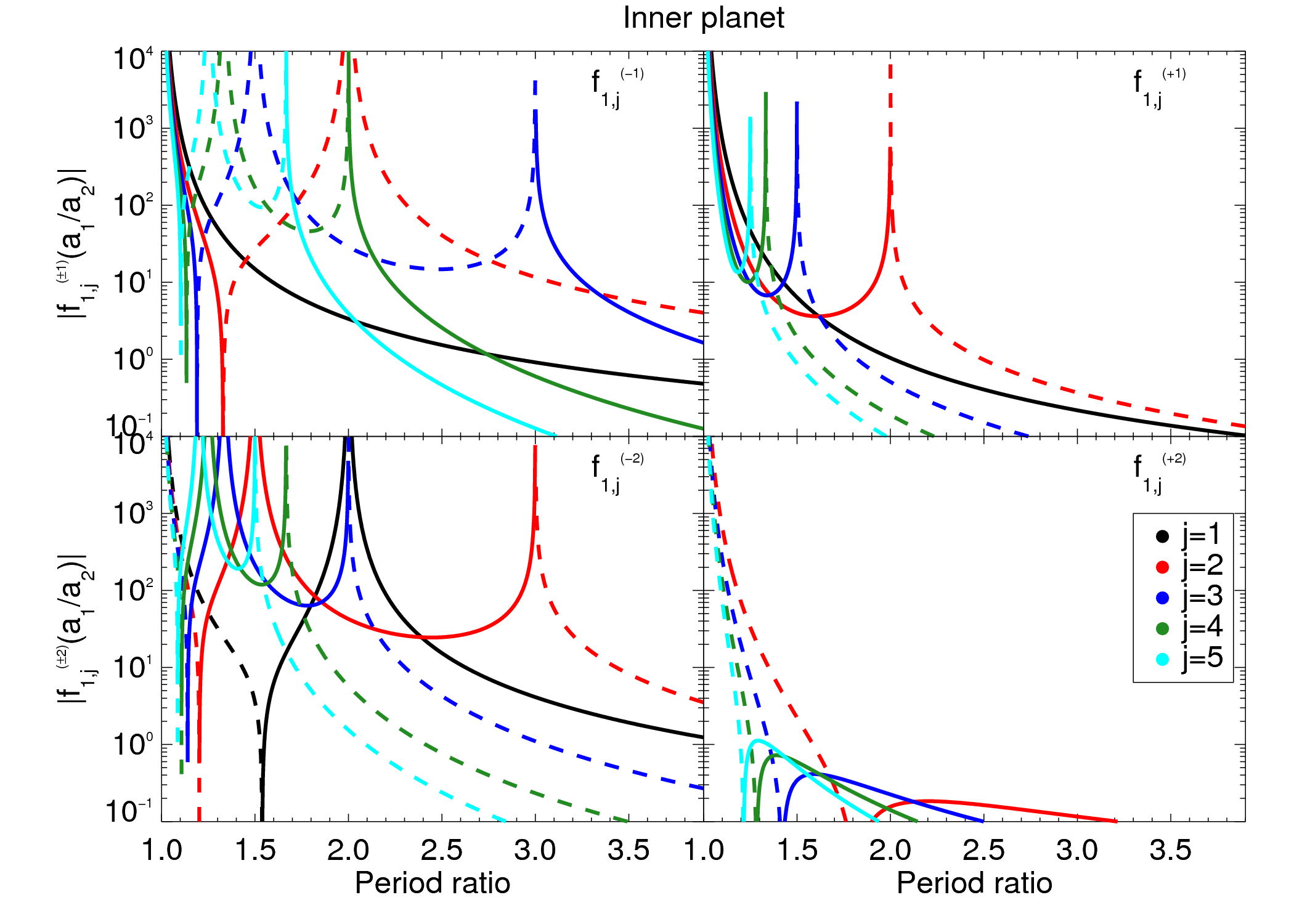}
\caption{Dimensionless coefficients, $f_{1,j}^{(\pm k)}$
for the inner planet.  The dashed lines show where the
coefficients are negative.}
\label{inner_coefficients}
\end{figure}

\begin{figure}
\centering
\includegraphics[width=0.8\hsize]{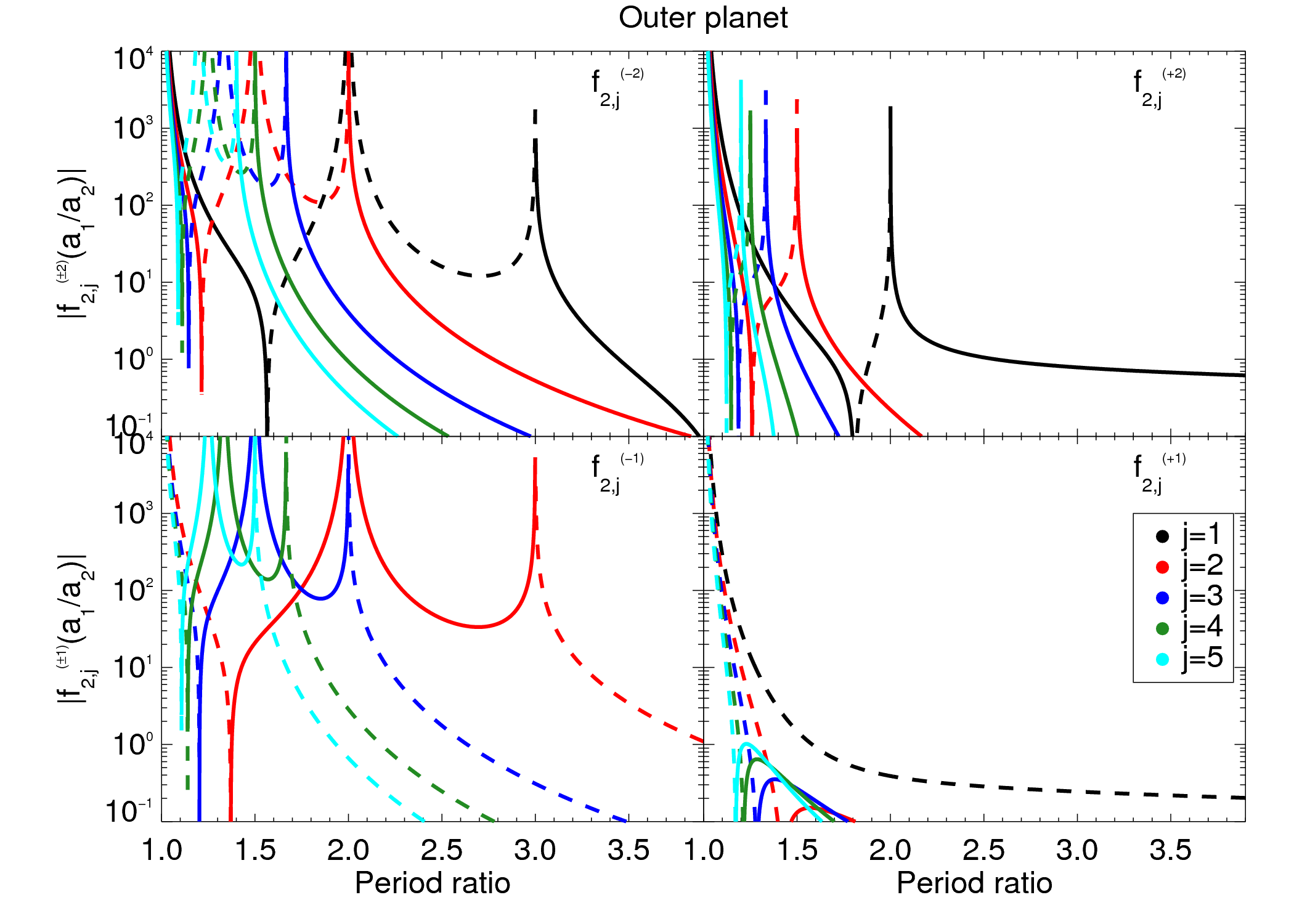}
\caption{Dimensionless coefficients, $f_{2,j}^{(\pm k)}$
for the outer planet. The dashed lines show where the
coefficients are negative.}
\label{outer_coefficients}
\end{figure}

In Figures (\ref{inner_coefficients}) and (\ref{outer_coefficients})
we plot the eccentricity-dependent coefficients, $f_{i,j}^{(\pm k)}(\alpha)$, 
as a function of period ratio, $P_2/P_1 \approx \alpha^{-3/2}$.  The zeroth-order eccentricity
coefficients are plotted in \citet{Deck2015}. The first-order coefficients
can show three singularities for the terms with superscripts
$(-1)$ and $(-2)$ near first-order resonance, second-order
resonance, and $\alpha=1$.

\subsection{Comparison with first order resonant equations}

LXW12 present a formula valid near (but not in)
first order mean-motion resonances that captures the behavior
of resonant terms in an elegant, but approximate, manner. Here
we compare the complete formulae given here to their near-resonant
formulae.

The expressions for $u$ and $v_{\pm}$ do not show the same
dependence in the denominator as the expressions in LXW12; their
expression just contains the resonant frequency, $jn_1-(j+1)n_2$, 
while ours contains additional frequencies.  We
carried out the partial
fraction expansion of $u$ to isolate the denominator which
matches LXW12's expression, and we
find that the expressions agree exactly with their expressions
(A28) and (A29).

To compare our full expression with LXW12's, we have computed the
eccentricity-dependent $j=2$ (near 2:3) expression for the inner 
and outer planets (as this term is unaffected by the indirect terms). We re-write the TTV formulae derived here and in LXW12 for the $i-$th planet perturbed by planet $k$ as 
\begin{align}
\delta t_i & = \frac{\mu_k}{n_i}\sum_{j\ge 1}\bigg[A_{i,i}^j e_i \sin{(j\lambda_k +\phi_{i,i}^j)}+A_{i,k}^j e_k \sin{(j\lambda_k +\phi_{i,k}^j)}\bigg]
\end{align}
where we have set $\theta_i = \lambda_i = 0$ at transit (this incurs some error, at order $e$, but that is a second order effect since we are comparing the TTV term linear in $e$). When written in this way, the amplitude and phase depend only on $\alpha,\varpi_1$, and $\varpi_2$.

\begin{figure}
\centering
\includegraphics[width=0.8\hsize]{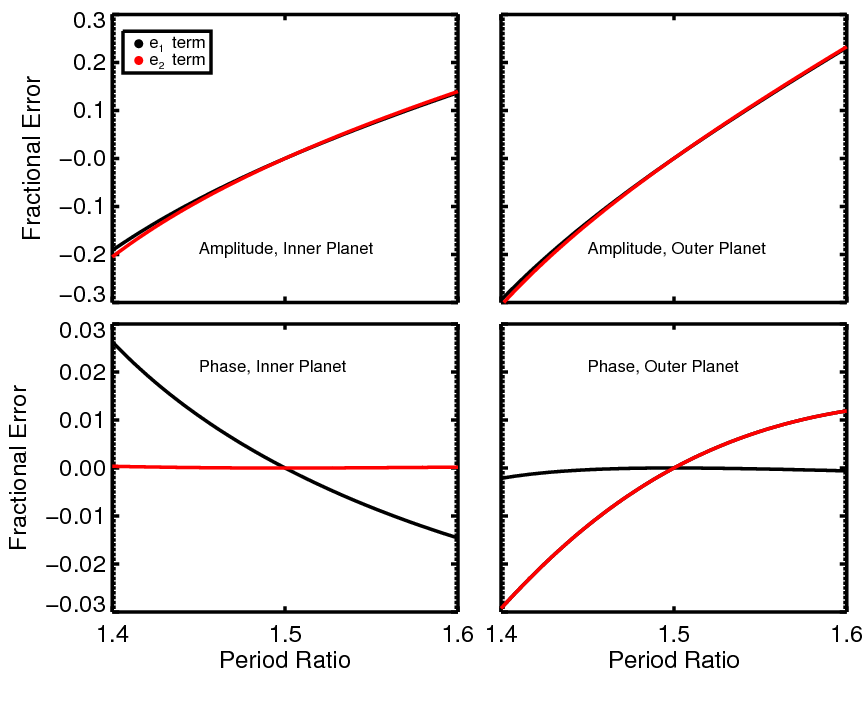}
\caption{Fractional error in the coefficients of the 3:2 resonant TTV expression given in LXW12 compared with the $j=2$ terms of our analytic expression.   Black indicates dependence
on the $e_1$ term, while red indicates the
dependence on the $e_2$ term.  The top show the fractional errors
in the amplitudes of the inner (left) and outer (right) planets.
The bottom shows the fractional error on the phases.}
\label{fig:lxw12}
\end{figure}

For the 3:2 resonance, the LXW12 resonant term depends on $e_1/(2n_1-3n_2)^2$ and
$e_2/(2n_1-3n_2)^2$, which both decline quickly away from resonance, and
thus other terms that depend on $e_1$ and $e_2$ make a more
significant contribution further from resonance; hence our formulae agree close to resonance but diverge away from exact commensurability.
Figure \ref{fig:lxw12} shows
the fractional error in the amplitude $A$ and phase $\phi$ of the terms that are proportional to the eccentricities of the planets for $\varpi_1\approx \varpi_2 =0.45$ radians.  The error in the LXW12 expression
(A28) and (A29)
reaches $\approx$20\% at a 5\% separation from exact resonance
in this case;  if we use the further approximate expression given
in their main paper in lieu of (A28), the discrepancy for the
inner planet increases to $\approx$30\%.  This is similar to the 
error in the zeroth-order
component of their expression \citep{Deck2015}.  
The zeroth- and first-order eccentricity terms have a different
dependence on longitudes:  for example, for the inner planet 
the zeroth order term scales as $e^{\mathbbm{i}(j+1)(\lambda_1-
\lambda_2)}$, while the first order term scales as $e^{\mathbbm{i}(j\lambda_1-(j+1)
\lambda_2)}$, where $\mathbbm{i}=\sqrt{-1}$.  Since the mean longitude of the inner planet is nearly identical at
each transit of the inner planet, the $\lambda_1$ term in the exponent is approximately 
constant, while the $\lambda_2$ dependence is identical in both 
terms, leading to aliasing of these coefficients.  Thus the error 
incurred in their approximation can lead to a different phase dependence
and amplitude for this aliased term away from resonance.

\subsection{Comparison with N-body integrations}

We have carried out extensive integrations of three-body
systems using {\it TTVFast} \citep{TTVFast}, and compared the results 
with the first-order analytic formulae (equation \ref{timing_solution}).
Note that the {\it TTVFast} code uses the convention of the
longitude of periastron being measured from the sky plane
to match the convention of radial velocity surveys, while
here we use the observer's line of sight as the reference
direction, as done in LXW12 and \citet{Deck2015}.  The
longitudes computed from the {\it TTVFast} code need to
have $\pi/2$ subtracted to make the plots shown below.
In addition, the orbital elements accepted by {\it TTVFast}
are the instantaneous/osculating orbital elements (initial
conditions) at the specified initial time, while the
orbital elements used in these formulae are the mean
orbital elements of the planets over the timescale of the
observations.

\subsubsection{Eccentricity and period ratio dependence}

\begin{figure}
\centering
\includegraphics[width=0.45\hsize]{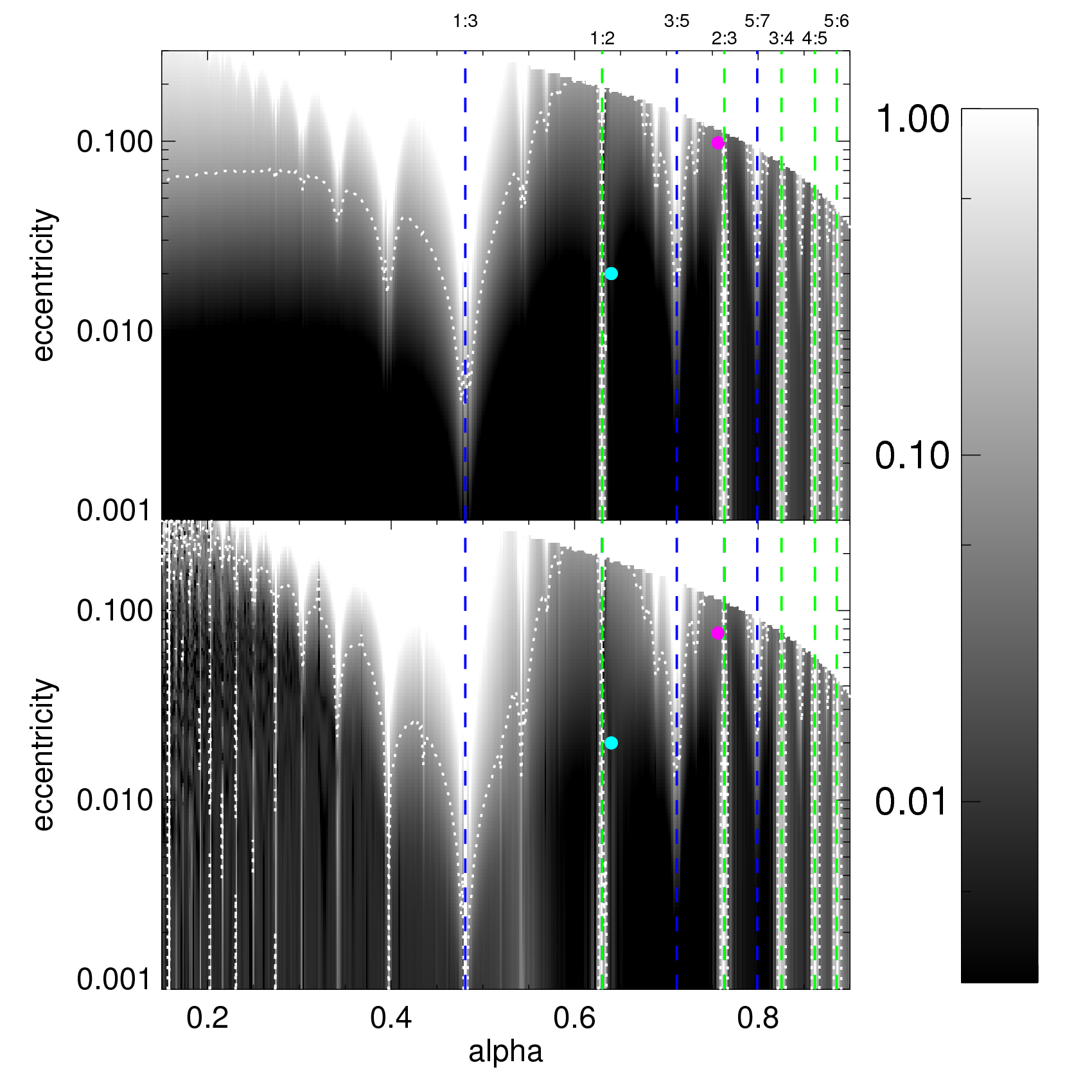}
\includegraphics[width=0.45\hsize]{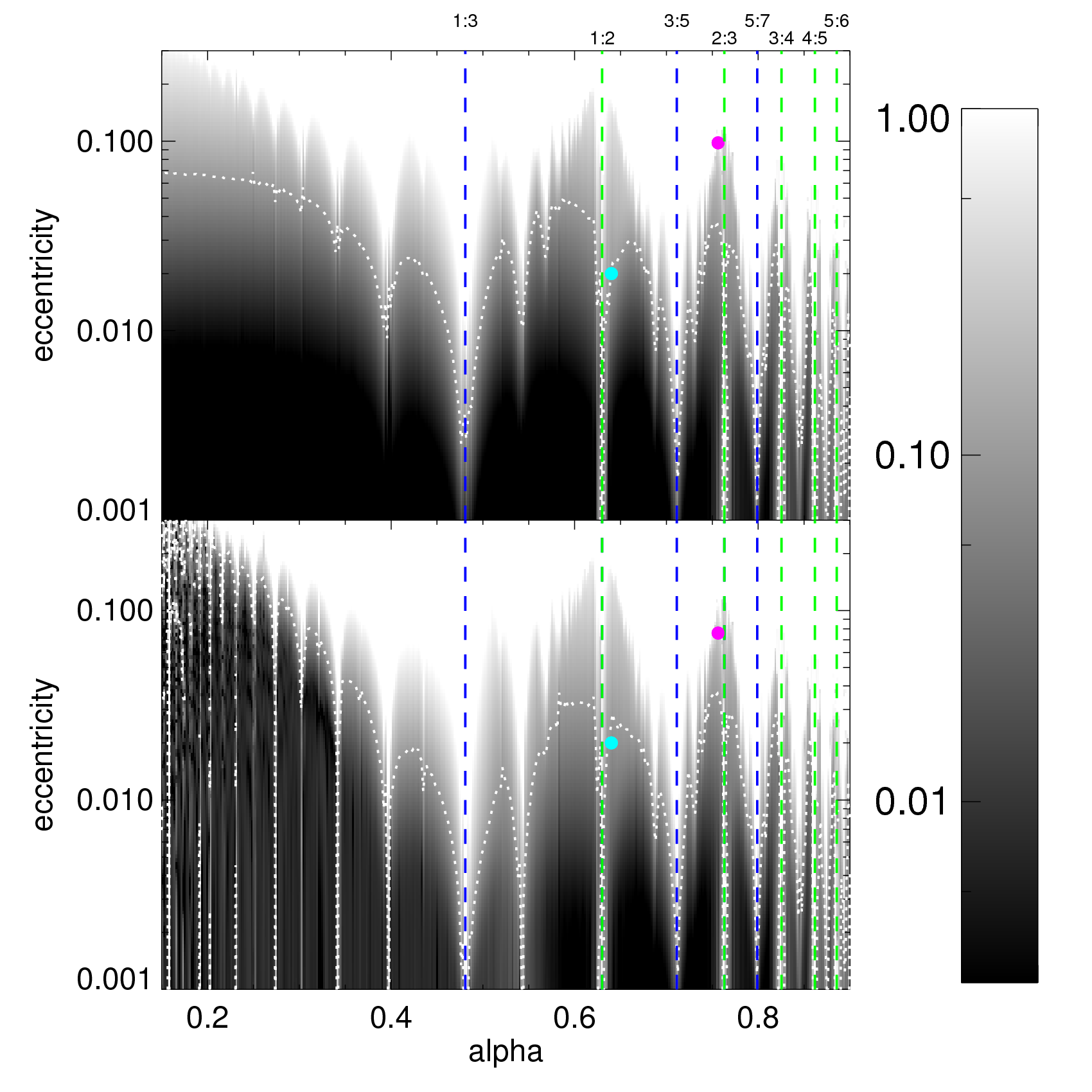}
\caption{Fractional precision of the analytic formula compared
with {\it TTVFast}. Left: Aligned longitudes of periastron ($\varpi_1=\varpi_2$);
Right: Anti-aligned longitudes of periastron ($\varpi_1=\varpi_2+\pi$).
The dotted lines indicate the 10\% precision
level.  Cyan dots show the approximate position of Kepler-18b/c at
the 85.15\% posterior eccentricity value, while magenta is used for
Kepler-28. The region in the upper right
is Hill unstable; these models were not computed, and default to 100\%
uncertainty in this plot.  The green dashed lines show the locations
of $j$:$j+1$ resonances, while the blue dashed lines show $j$:$j+2$.}
\label{fig01}
\end{figure}

Figure \ref{fig01} compares the precision of the analytic
formula as a function
of $\alpha = (P_1/P_2)^{2/3}$ and eccentricity of both planets,
which are set to be equal, $e_1=e_2$.  We have set $\varpi_1 = \varpi_2+\pi$,
which we found (approximately) maximizes the discrepancy of the analytic model
compared with the N-body model, and $\varpi_1=\varpi_2$ which
(approximately) minimizes the discrepancy; hence the figures bracket 
the precision  of the analytic model.  This is due to the fact that
the anti-aligned longitude geometry causes the planets to be closer
at conjunctions that occur when the inner planet is at apoapse and
the outer is at periapse;  their proximity at these conjunctions causes their 
gravitational interactions to be more sensitive to deviations from the 
epicyclic approximation, which are second order in eccentricity, and
thus missing from our computation. We have assumed
that the period of the inner planet is $P_1=30$ days, and we have
integrated the system with {\it TTVFast} for 1600 days, about the
duration of the initial Kepler mission, assuming plane-parallel orbits.  
For these tests we assume $\mu_1 = \mu_2 = 10^{-5}$, and we
selected random values for the longitudes of the planets at the initial time.  
For each set of initial conditions, we output the
orbital elements at regular intervals during the N-body integration,
from which we computed the average orbital elements over the duration
of the integration.  These averaged orbital elements were used
for computing the amplitudes of the analytic model, which we
summed up to $j_{max}=10$.  We optimized
the fit of the analytic formula to the numerical TTVs by allowing
the ephemerides of the planets to vary in the formula, but holding the eccentricity
vectors and mass ratios fixed at the values computed from the
time-averaged N-body simulation, while  we computed $\alpha$ in the
analytic formula from the
ratio of the periods derived from the best-fit ephemerides.

The fractional precision was computed from the RMS of
the residuals of the best analytic model fit to the TTVs,
divided by the RMS of the TTVs computed from the N-body integration.
Figure \ref{fig01} shows that the formula works to better than
10\% precision for a wide range of $\alpha-e$ parameter space.
However, it fails near resonances, most significantly for
the $j$:$j+1$ resonances indicated in green, and $j$:$j+2$
in blue. For the outer planet, there are narrow regions
near $1$:$j$ period ratios for which the formula does poorly.
The disagreement grows in breadth for larger eccentricities.
This diagram can be used to pinpoint the relevance of the
analytic formula for a particular system, and we suggest that
the analytic formulae should be used with caution in the regions
where the formula disagrees by more than 10\% precision. 

Most of the regions where the formula fails are near resonance. 
In these cases, the residuals can frequently be fit by sinusoidal variations 
at the relevant resonant frequencies of the higher order
resonant terms that are not captured in the first-order model;
when including these sinusoidal terms in the fit, the residuals
drop dramatically near the resonances.  Thus, the analytic first-order
solution plus a sinusoid with arbitrary amplitude and phase
can be used for systems in which only the shape of the transit timing variations 
plus the specific variations of the non-resonant terms is
necessary (although this approach breaks down for large
enough eccentricity).

\subsubsection{Mass dependence}

We have carried out simulations for a range of masses, keeping
$m_1=m_2 (\mu_1=\mu_2)$.  We find that the fractional error of the analytic
formula grows near resonances and near $\alpha=1$ as the mass 
increases with weaker dependence upon eccentricity.  For small 
$\alpha$, the formula works well up to $m_1 = m_2 = 10^{-3} m_\star$, 
while near $1$:$2$ period ratio, for example, the error broadens 
around the resonance.

\begin{figure}
\centering
\includegraphics[width=0.5\hsize]{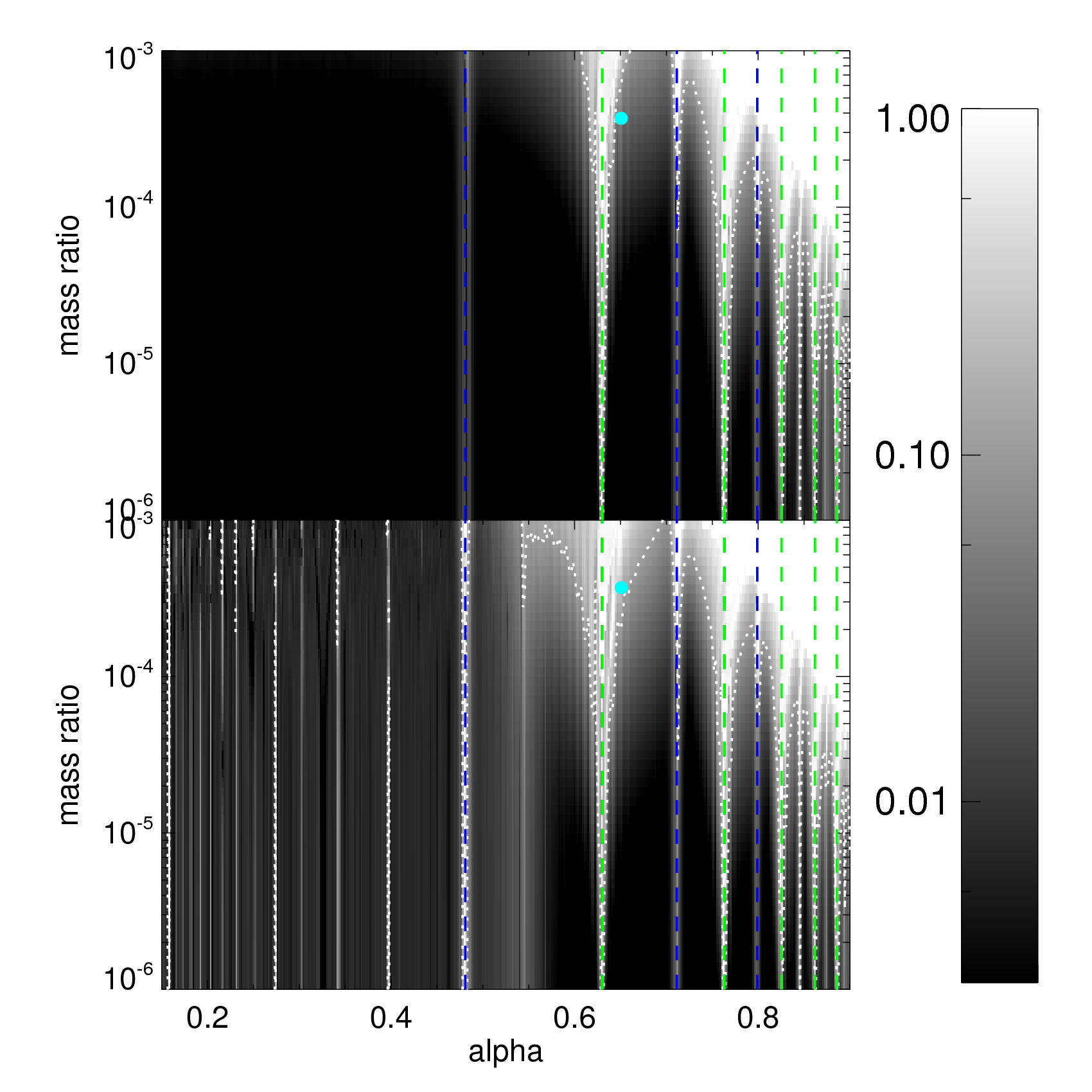}
\caption{Fractional precision of the analytic formula compared
with {\it TTVFast} versus $\alpha$ and mass ratio of the planets
to the star ($m_1=m_2$).   Top:  inner planet; bottom: outer planet.
The dotted line indicates the 10\% precision
level.  Cyan dots show the approximate position of PH3 (although
note that PH3 does not have equal masses; however, the plot is
similar for parameters appropriate for PH3). The region 
in the upper right
is Hill unstable; these models were not computed, and default to 100\%
uncertainty in this plot.  The green dashed lines show the locations
of $j$:$j+1$ resonances, while the blue dashed lines show $j$:$j+2$.}
\label{mass_error}
\end{figure}

Figure \ref{mass_error} shows the fractional error in the
formula (computed as in the eccentricity dependent case)
for $e_1 = e_2 =0.001$ (the mass dependence of the precision
is nearly independent of eccentricity) with $\varpi_1=\varpi_2+\pi$
(the results look very similar for $\varpi_1=\varpi_2$).
The formula is accurate for a broad range of masses, but 
for some systems, such as Planet Hunters 3c/d, indicated
in Cyan in Fig. \ref{mass_error}, the discrepancy becomes
large, $\approx 10$\% (the masses and eccentricity vectors
for this plot differ from PH3c/d, but a plot made for the
parameters of that pair of planets looks very similar).

\subsection{Comparison with two-planet systems}

We have carried out fits to systems with two interacting planets
described in LXW12, and we have re-fit Planet Hunters (PH3) c/d.  
We have carried out N-body dynamical analyses using
{\it TTVFast} in addition to fits with the analytic
first-order formula in order to assess its utility in analyzing
multi-planet systems.

\begin{figure}
\centering
\includegraphics[width=0.45\hsize]{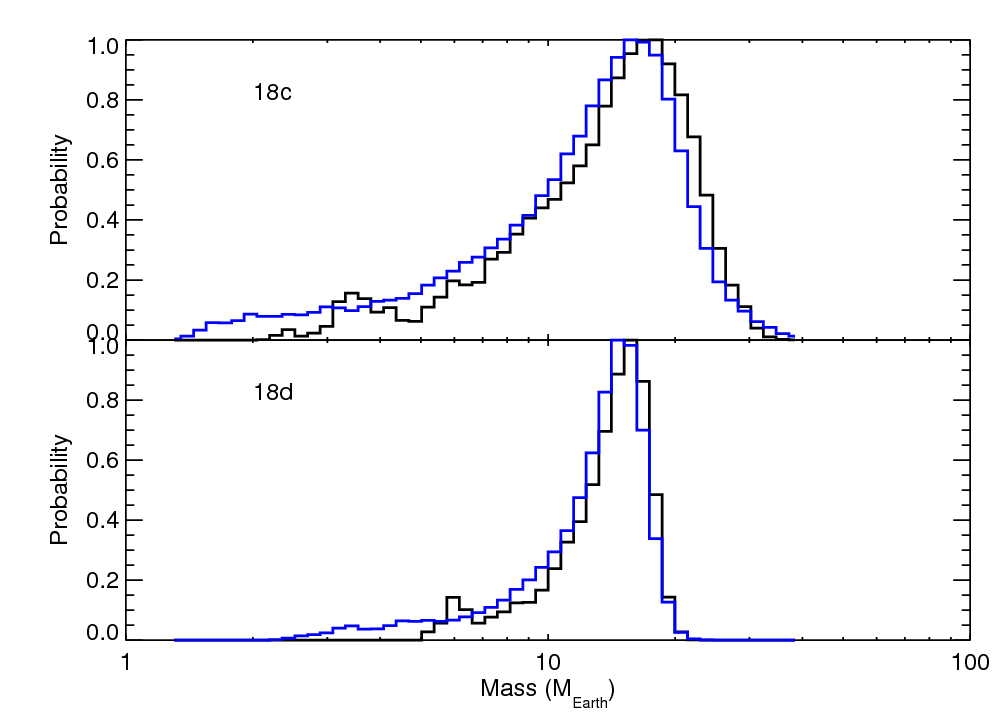}
\includegraphics[width=0.45\hsize]{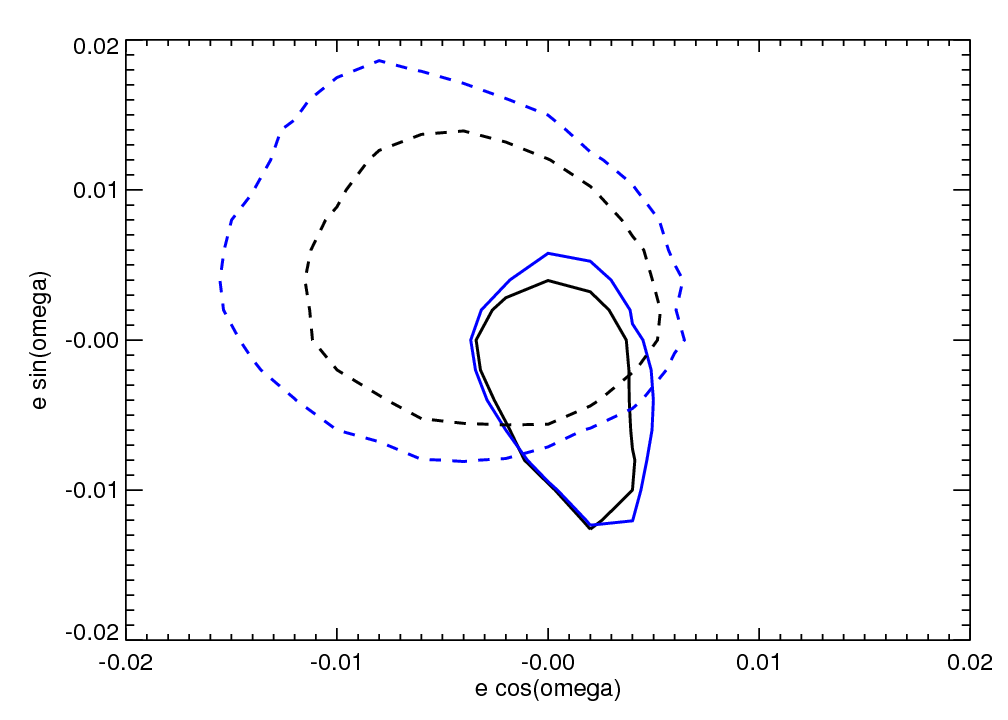}
\caption{Comparison of numerical and analytic analysis of the
transit times of Kepler-18.  Left: histogram of the masses
of each planet.  Right: comparison of the 68\% confidence
distribution of the eccentricity vectors of the two planets 
(18c solid; 18d dashed).
In each case black indicates the results from the N-body
analysis, while blue indicates the results of the analytic
formula.}
\label{fig02}
\end{figure}

Our first case is Kepler-18c/d, which was published by \citet{Cochran2011}
and also analyzed by LXW12.  We used the same transit times
and uncertainties from the \citet{Cochran2011} paper to
allow for direct comparison to their results;  these transit
times were also used in LXW12. We carried out a markov chain monte
carlo (MCMC) analysis using an affine-invariant population approach
\citep{Goodman2010}.  We allow a multiplicative factor for the
timing uncertainties on each planet and we place a uniform
prior on the eccentricity of each planet \citep{Ford2006}.
Figure \ref{fig02} shows a comparison of the results of the
MCMC analyses with N-body integration versus the analytic formulae
with $j_{max}=5$.
For this system the mass ratios are $\approx 5 \times 10^{-5}$
and the eccentricities are of order $0-0.02$ (1-$\sigma$), while $\alpha
\approx 0.64$, which is a regime in which the first-order
formula is accurate to $<$9\% compared with N-body integration
(see Figure \ref{fig01}; the cyan dot indicates the approximate
location of the upper end of the eccentricities of Kepler-18c/d).
For conversion of the mass ratios to planet masses, we assume
that $m_\star = 0.972 M_\odot$ (we ignore the uncertainty on
this stellar mass).
The masses of the planets derived from N-body are 
$m_1 (nbody)=  14.7_{-6.7}^{+5.4} M_\oplus$ and 
$m_2 (nbody) = 14.3_{-4.1}^{+2.3} M_\oplus$, while from the
analytic formula are $m_1 (analytic) =   13.2_{-  7.0}^{+  5.4}
M_\oplus$ and $m_2 (analytic) =   13.6_{-  4.9}^{+  2.4} M_\oplus$.
These are well within $1-\sigma$ of one another, the markov
chain posteriors show very similar distributions (Fig. \ref{fig02}).  
We note that these results differ from those reported by \citet{Cochran2011}
who used N-body to estimate the transit times, found the
best fit using Levenberg-Marquardt optimization, and estimated 
the uncertainties from the Hessian matrix at the best fit
parameters rather than a full posterior analysis.  Our results 
also differ from the estimates in LXW12, which only solve for a `nominal'
mass assuming zero eccentricity using the approximate near-resonant
formula.  We feel that our results should be superior to these
prior results, and warrant a more extensive analysis with the
full Kepler dataset as well as radial velocity measurements.

\begin{figure}
\centering
\includegraphics[width=0.45\hsize]{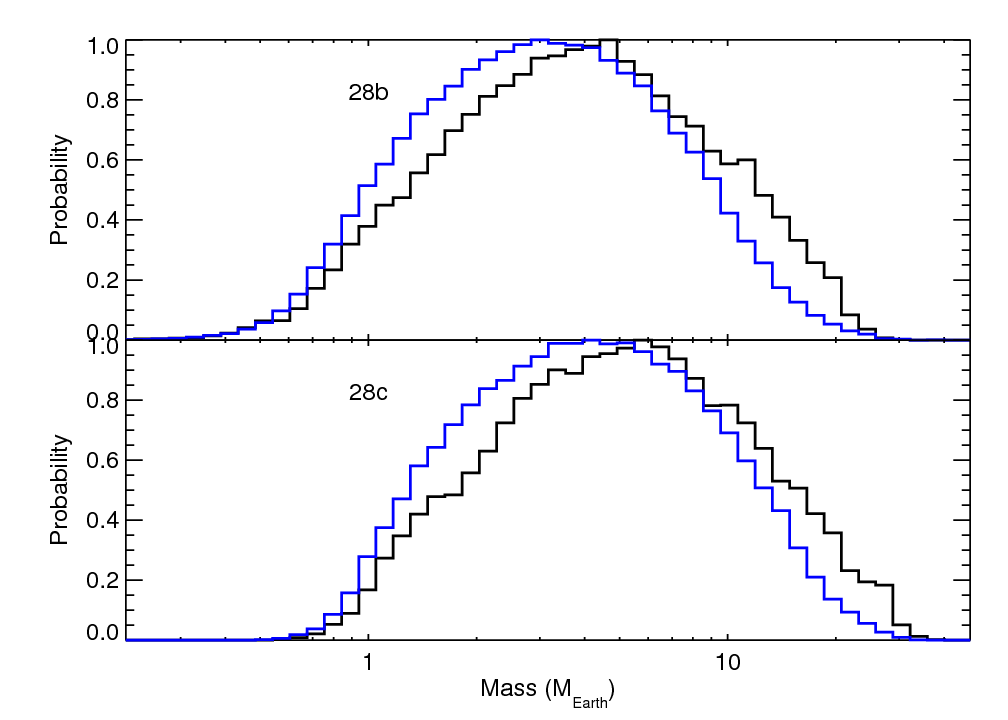}
\includegraphics[width=0.45\hsize]{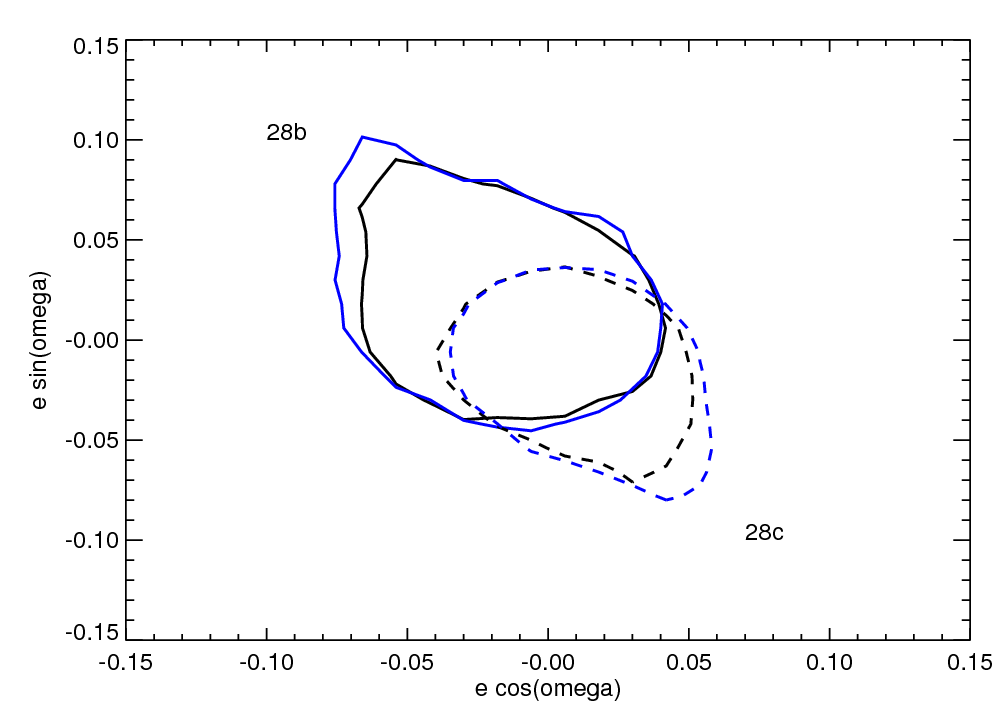}
\caption{Comparison of numerical and analytic analysis of the
transit times of Kepler-28.  Left: histogram of the masses
of each planet.  Right: comparison of the 68\% confidence
distribution of the eccentricity vectors of the two planets 
(28b solid; 28c dashed).
In each case black indicates the results from the N-body
analysis, while blue indicates the results of the analytic
formula.}
\label{Kepler28}
\end{figure}

We next compared analyses of the Kepler-28 system, which
was originally studied by \citet{Steffen2012}, and included in the analysis
of LXW12. We used the transit times published by \citet{Steffen2012},
and followed the same procedure as Kepler-18.
Figure \ref{Kepler28} shows a comparison of the masses and
eccentricity vectors for this system, which has
a mean period ratio of $\langle P_2/P_1 \rangle = 1.52$, just
wide of 2:3 period ratio, corresponding to $\alpha=0.7563$.
The masses are poorly constrained due to
the degeneracy with eccentricity, which allows the eccentricity
to wander to larger values.  
For conversion of the mass ratios to planet masses, we assume
that $m_\star = 0.89 M_\odot$ (we ignore the uncertainty on
this stellar mass).  A comparison between the mass
constraints from N-body and the analytic formula gives:
$m_1 (nbody)= 3.8_{-2.3}^{+4.6} M_\oplus$ versus
$m_1 (analytic)=  3.1_{-1.7}^{+3.4} M_\oplus$, and
$m_2 (nbody)= 5.1_{-3.0}^{+5.9} M_\oplus$ versus
$m_2 (analytic)=  4.1_{-2.3}^{+4.6} M_\oplus$.  The 85\% confidence
value of $e_1$ is 0.098, while for $e_2$ is 0.076, while
the longitudes of periastron within the posterior distribution
are primarily anti-aligned;
the location of these points is
indicated with a magenta datapoint in Figure \ref{fig01}.
Note that in the anti-aligned $\varpi$ case the analytic
formula is valid to larger eccentricities, and thus is
adequate to describe this system.

\begin{figure}
\centering
\includegraphics[width=0.45\hsize]{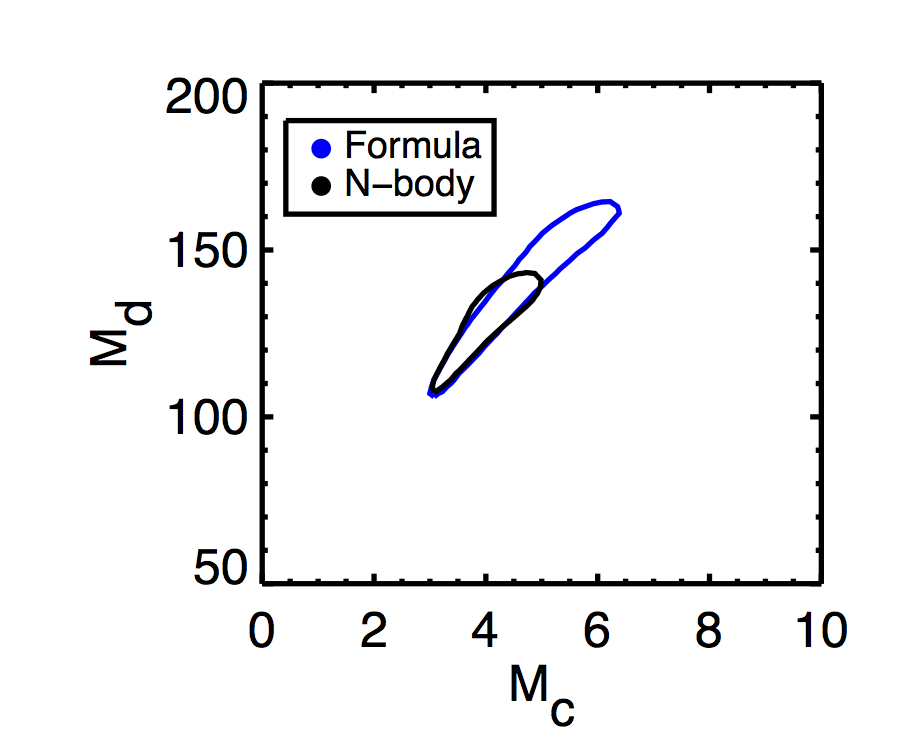} 
\includegraphics[width=0.45\hsize]{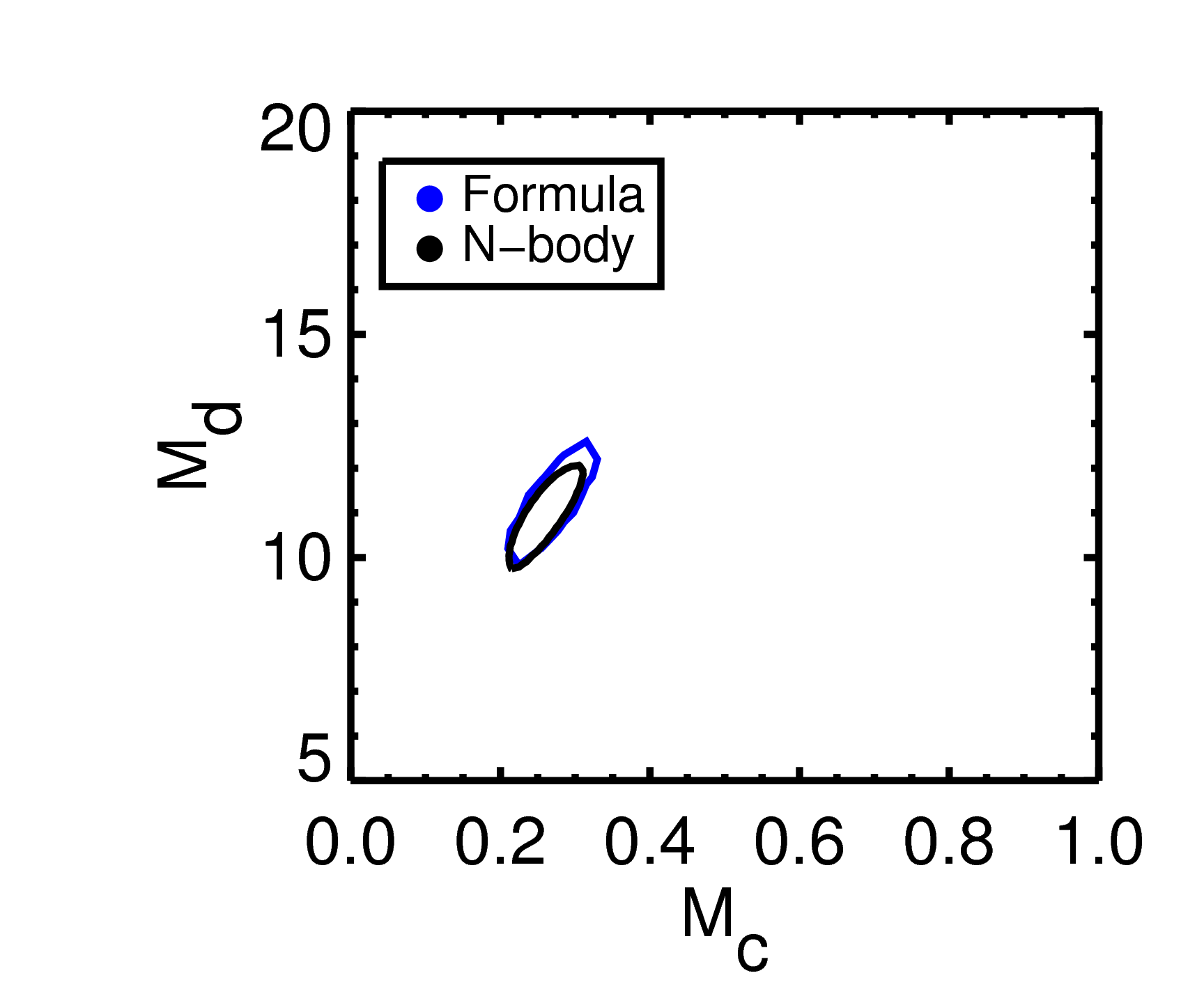}
\caption{Comparison of numerical and analytic analysis of the
transit times of Planet Hunters 3c/d.  Left: 68\%
confidence contour of the masses
of both planets.  Right: comparison of the 68\% confidence
limits with the mass of the both planets reduced by 
a factor of 10.}
\label{ph3_masses}
\end{figure}

As Figure \ref{mass_error} indicates, the Planet Hunters 3 (PH3)
system, the outer two planets (c/d) analyzed in \citet{Deck2015},
has a large discrepancy due to the large mass of the outer
planet.  The excellent agreement with the chopping formula
given in \citet{Deck2015} is still imperfect;  the figure
in that paper was mistakenly produced with larger timing error bars than used
in \citet{Schmitt2014} which caused the agreement to appear
slightly better than the first-order formula indicates.
 Figure \ref{ph3_masses} shows the
results of a comparison of N-body and analytic fits to
the PH3 transit times given in \citet{Schmitt2014}; the planet
masses assume $m_\star = 1M_\odot$.  The
analytic formula gives a significant discrepancy due to
the large mass of the outer planet and due to the proximity to
the 1:2 period ratio.

To confirm that the large mass of PH3d led to this discrepancy, 
we took the best-fit parameters resulting from our N-body analysis 
of the real PH3 data, and reduced the masses of the outer two planets 
by a factor of 10. Using the outer two planets alone, we simulated 
transit times and added Gaussian noise at a level of $1/10$ 
that of the noise of the real data to maintain the same signal-to-noise
ratio as the actual data. We then modeled this simulated 
data using {\it TTVFast} and the analytic formulae.  We found that the
agreement between the N-body and analytic analyses becomes
excellent (Fig.\ \ref{ph3_masses}).

\subsection{Comparison with multi-planet systems}

The two-body solution can be used for more than two bodies by addition
of two-body TTV solutions for each pair of two planets \citep{Lissauer2011}:
\begin{equation}
\delta t_{i_1} = \sum_{i_2 \ne i_1} \delta t_{i_1,i_2},
\end{equation}
where $\delta t_{i_1,i_2}$ are the solutions from equation (\ref{timing_solution})
for the $i_1$th planet due to the $i_2$th planet.  The sum over $j$ for
each pair of planets can be carried up to $j_{max}$ to give sufficient
precision for that pair of planets that is smaller than the measured
timing precision.

\begin{figure}
\centering
\includegraphics[width=0.5\hsize]{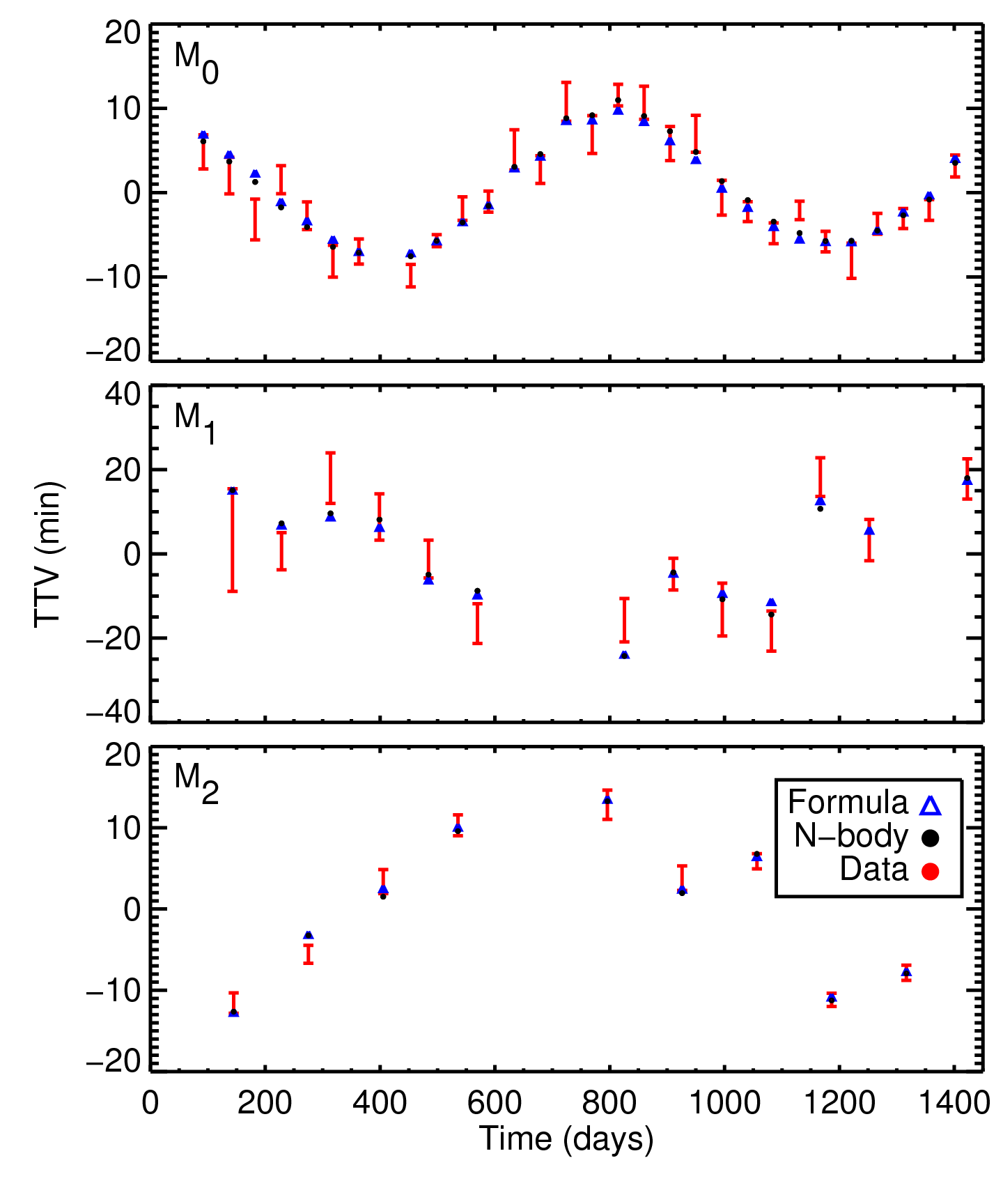}
\caption{Transit timing variations from \citet{Masuda} 
for Kepler-51 (red; 0, 1, 2 stand for 51b,c, 620.02),
compared with the best-fit N-body model computed with
{\it TTVFast} (black), and the best-fit two-planet, first-order
eccentricity formula summed over pairs of planets (blue).}
\label{Kep51_TTVs}
\end{figure}

\begin{figure}
\centering
\includegraphics[width=0.35\hsize]{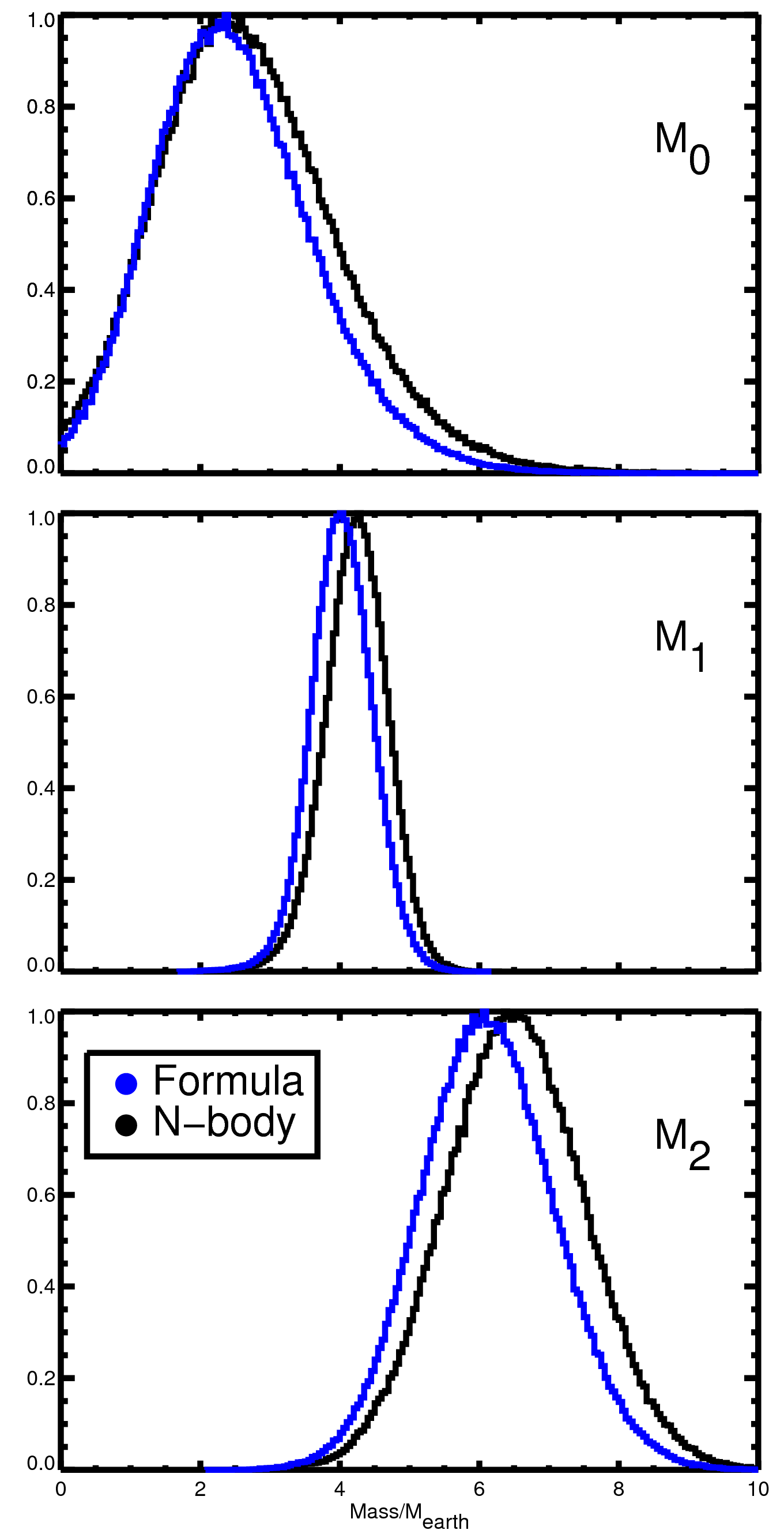}
\includegraphics[width=0.35\hsize]{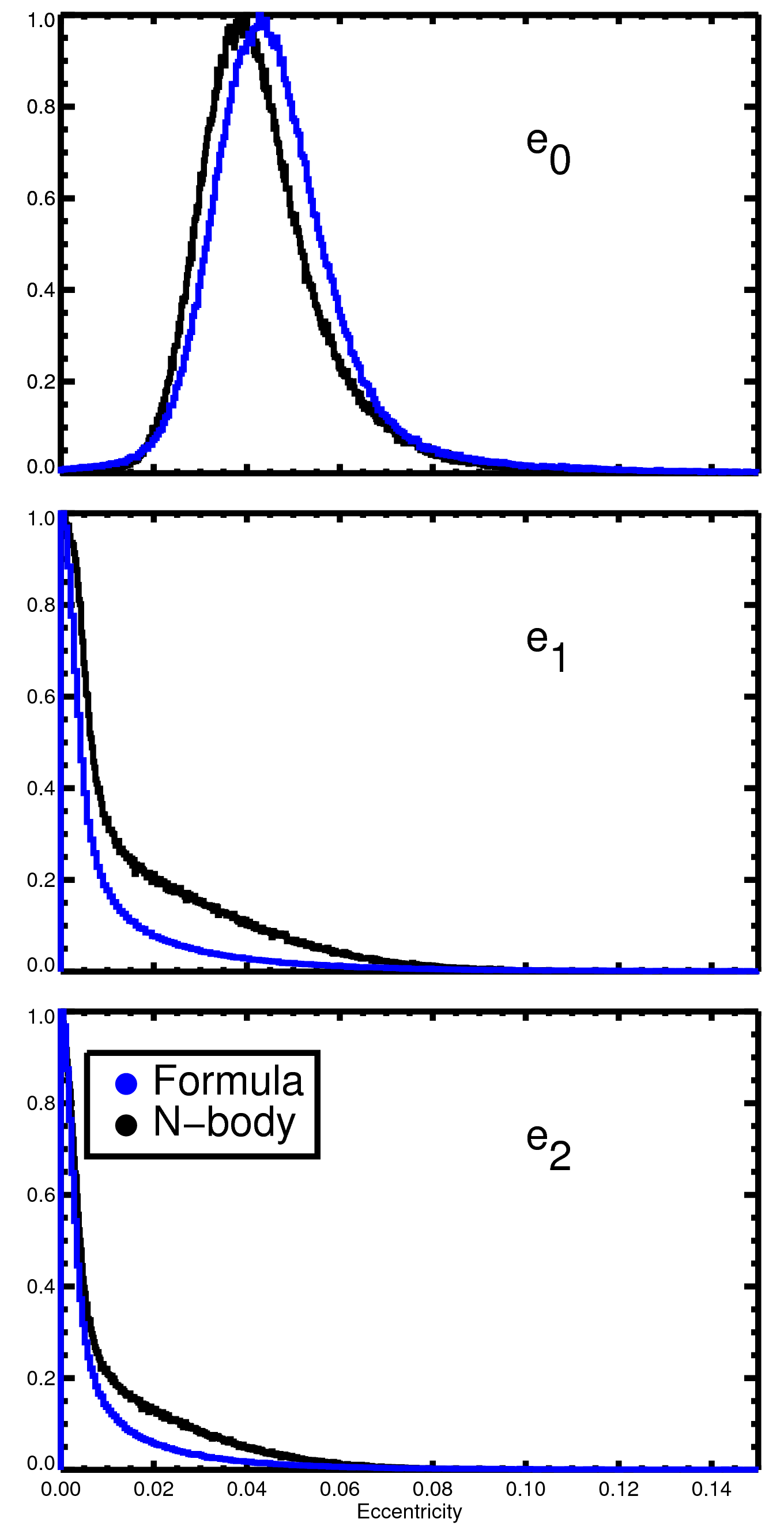}
\caption{Comparison of numerical and analytic analyses of the
transit times of Kepler-51 (0, 1, 2 stand for 51b,c, 620.02).  
Left: histograms of the masses of each planet.  Right: 
histograms of the eccentricities of the planets.}
\label{Kepler51}
\end{figure}

Our first system of study is Kepler-51 \citep{Masuda}, consisting of
planets with period ratios close to 1:2:3.  We used the transit times 
reported in \citet{Masuda} to carry out dynamical models with N-body/{\it TTVFast} and with
an analytic TTV signal given as the sum of the TTVs of the three adjacent 
pairs of planets. We included up to $j=6$ in the TTV signals.   The results show
excellent agreement;  Figure \ref{Kep51_TTVs} shows the measured
transit-timing variations, as well as the best-fit N-body and
analytic TTVs.  The results of the MCMC analyses are compared in
Figure \ref{Kepler51} which shows the posterior distribution of
masses and eccentricities measured with both analyses, assuming
$m_\star = 1 M_\odot$.  For two
of the planets the eccentricities are consistent with zero;  in these
cases the tail of the eccentricity histogram is heavier for the
N-body than the for the analytic formula.  The masses of the inner
two planets from our N-body and analytic MCMC analyses agree well 
with the masses from the analysis in \citet{Masuda}, while the
mass of the outer planet is small by $\approx1-\sigma$.

\section{Numerical implementation and speed}\label{numerics}

The primary computational burden of equation (\ref{timing_solution}) lies
in computing the Laplace coefficients.  We use a series solution
for these coefficients, which gives both speed and accuracy,
using code shared by Jack Wisdom.  The secondary computational burden is
in computing the sine and cosine terms, which involves four
angles, and thus requires eight evaluations.  We carry
out the computation of higher $j$ sines and cosines using
trigonometric addition formulae, which means we only need
to compute eight trigonometric functions at each transit time
once;  the rest are gotten from addition and multiplication of
these.

In the cases that we run a Markov chain for a set of planets, the initial value of
$\alpha$ is known fairly well from the period ratio of the planets.
In this case the Laplace coefficients and their derivatives needed
for the solution can be Taylor expanded at the $\alpha$ given by
an initial fit to the transit times, and these coefficients can be 
stored for evaluation of the coefficients at slightly different values of
$\alpha$ encountered during the MCMC simulation.  This approach
would not work if $\alpha$ is being varied over a grid (for example,
in the case of searching for a perturbing planet with unknown
period); however, computational efficiency can still be achieved
by reusing the Laplace coefficients at different eccentricities 
\citep{NesvornyMorbidelli2008}.

We have coded the first-order formula, equation (\ref{timing_solution}),
in C, IDL, Python, and Julia \citep{Julia}.  We carried out a benchmark comparison of
the Julia implementation of the formula with the C implementation
of {\it TTVFast}, and we find that it is $400\times$ faster when the 
Laplace coefficients are approximated from a Taylor expansion.  As
{\it TTVFast} is about 20 times faster than TTVs computed with
standard N-body integrators, this represents nearly four orders
of magnitude in speed up, similar to that found by
\citet{NesvornyMorbidelli2008}.  Note that if integer
period ratios are chosen, sometimes the denominators of $u$ and
$v_\pm$ can become infinite, causing divergence;  we expect that
this will not be encountered in practice as the formulae only
apply to non-resonant planets.

The code implementing these equations may be accessed at
\url{https://github.com/ericagol/TTVFaster}.

\section{Discussion and conclusions}\label{conclusions}

In modeling transit timing variations, degeneracies and computational
speed can each prohibit the accurate measurement of transiting planet
masses and orbital properties, with their attendant uncertainties.
The degeneracy due to aliasing near first order resonances (LXW12)
can be broken with very high signal-to-noise due to the slight
difference in the eccentricity dependence as a function of period
ratio, as well the presence of perturbations at other frequencies
\citep{Deck2015}.  Here we have tried to improve the modeling of
the terms which are linearly dependent upon eccentricity to provide
a higher-fidelity analytic model to address both the degeneracy and
the computation barriers.  

To this end, we have presented a first-order solution in eccentricity
and mass ratio to the plane-parallel, near-circular 3-body problem 
on timescales shorter than the secular timescale.  This improves to 
first order in eccentricity the original solution given in \citet{Agol2005} 
which was derived to zeroth order in eccentricity and first order in 
mass ratio (this solution has also been given in different forms in
\citealt{Nesvorny2014} and \citealt{Deck2015}).  The expressions are accurate compared with 
numerical integration over a wide range of parameter space relevant
to the hundreds of multi-transiting planetary systems being found at short 
orbital periods with Kepler \citep{Rowe2014,Lissauer2014}.  We find that
this expression is more accurate than the stripped-down near-resonant
formula given in LXW12, although their formula has a simpler form
which clearly highlights the mass-eccentricity degeneracy.  The first-order
eccentricity formulae also 
can be used to model more than two planets with linear combinations of 
2-planet formulae, and works well for the system we tested here, Kepler-51.

We used an approach starting with the Newtonian
equations of motion rather than the Hamiltonian, and compute the
perturbed polar coordinates of the planets' orbits;  this approach
is akin to solving dispersion relations of differential equations
for mode and stability analysis (for example, the magneto-rotational
instability is derived with this approach, \citealt{Chandra1961,Balbus1998}).  
The unperturbed solution to these differential equations
represents Keplerian motion expanded in eccentricity.
The terms with various frequencies in the disturbing function give
an inhomogeneous component to the solution, which cause TTVs to
vary at frequencies which depend upon integer combinations of the 
orbital frequencies of the two planets.  Since the answer obtained
in the end is the same as in the Hamiltonian (for ${\cal O}(e^0)$) 
and canonical transformation approaches, this approach
might be useful pedagogically for those more familiar with
stability analyses.  In addition, this approach might be useful
for other problems, such as a stability analysis of a two-planet system
or for carrying out the TTV computation to second-order in mass 
ratios $\mu_i=m_i/m_\star$. The latter is interesting as it would reveal 
how TTVs of a transiting planet may be used to measure the mass of 
that planet (and not just the mass of the perturbing planet).

We expect that these formulae will be used in carrying out initial fits
to multi-transiting planets which show TTVs \citep{Tsevi}, in searching for companion
perturbing planets to isolated planets showing TTVs, in characterizing
multi-planet systems with TTVs to confirm and check for convergence
of N-body MCMC analyses, in forecasting TTV amplitude for follow-up
measurement, in estimating the optimum times for transit observation, and in
making predictions for transit times to plan observations.  It should be
useful for estimating the densities, masses, and radii of the host stars and their exoplanets
\citep{Agol2005,Montet2012,Hadden2014,Kipping2014,JontofHutter2015}, 
and in comparing the TTV solutions to radial velocity solutions for the masses 
and orbits of exoplanets. The analytic 
nature of our solution should be amenable to automatic differentiation 
\citep{Fournier2012}, which could speed up optimization based on gradient 
computation, and could also enable Hamiltonian/Hybrid Markov Chain Monte 
Carlo \citep{neal2011mcmc}.  When a large number of planets transit a
star and each show evidence for dynamical interactions, the number of
free parameters describing the system becomes large, and thus MCMC
becomes prohibitively computationally expensive.  The first-order
analytic formula developed here can be used for modeling these systems
if their masses and eccentricities are in the allowable range.  As
the formula is about 400 times faster than {\it TTVFast}, which is
already about 20 times faster than Bulirsch-Stoer based integrators,
the total speedup of about 8000 should make running chains long enough
to converge more feasible, especially in tandem with parallel computation
which can be easily adapted for population MCMC \citep{ForemanMackey2013}.  
The analytic formula
also has the advantage of being able to pinpoint which features in the
TTVs constrain the parameters of the system (LWX12).  Since the TTVs
of a planet display harmonics of the perturbing planet \citep{Deck2015},
the amplitudes and phases of each of these harmonics can be measured
directly from the TTVs, and then these can be used to place individual
constraints upon the masses and eccentricity vectors of the planets.
The regions where the constraints overlap may reveal the consistency
and uniqueness of the solution for the system parameters in some
cases \citep{Deck2015}.

Although in principle TTVs allow for unique measurements of planet mass 
and eccentricities, degeneracies between these parameters are often found 
for systems with low signal to noise. In these cases, the eccentricities 
can become extremely large, indicating unstable orbits, as long as the 
masses are adjusted in a corresponding manner. We applied Hill stability 
to avoid this problem when using the analytic formulae; the
full N-body computation avoids this issue naturally since large eccentricities
introduce second-order (and higher) variations (that our calculation
ignores) which prevent the high-eccentricity cases from fitting the
data well.  It may be possible to break some degeneracies with transit
duration variations \citep{Pl2008,Nesvorn2013}, which can be computed
with the same formalism we have described here, albeit in the plane-parallel
limit.

The first-order formulae described here could be extended to higher order 
in eccentricity and/or mass ratio,  albeit with much more computational 
effort.  A slightly more accurate formula might be obtained by computing
the longitudes from Kepler's equation at the times of transit of each
planet rather than using the first-order eccentricity formula (\ref{longitude}),
as well as using the exact formula for $\dot\theta_{i}$ at the transit
times in equation (\ref{ttv});  this requires very little additional 
computational effort, but $\delta \theta_i$ will still be only accurate
to first order in eccentricity.
The solutions for the perturbed polar coordinates, $(\delta 
r_i, \delta\theta_i)$, can be derived in the same manner that we have 
derived the transit timing variations.  These are needed for carrying out
the higher-order perturbation solutions, and in turn could be used for 
modeling astrometric variations, radial velocity varations, and pulsar 
timing variations of host stars to account for the interactions of planets to 
first order in eccentricity.

\acknowledgements

EA acknowledges support from NASA grants NNX13AF20G, NNX13AF62G, and 
NASA Astrobiology Institute’s Virtual Planetary Laboratory, supported 
by NASA under cooperative agreement NNH05ZDA001C. This research was 
supported in part by the National Science Foundation under Grant No. 
NSF PHY11-25915. EA thanks the Kavli Institute for Theoretical
Physics and the organizers of the ``Dynamics and Evolution of Earth-like Planets"
workshop where a portion of this work was completed;  this manuscript
is preprint number NSF-KITP-15-132. KD acknowledges
support from the Joint Center for Planetary Astronomy fellowship.  We 
thank Jack Wisdom for sharing laplace.c which
computes Laplace coefficients and their derivatives with series summation, we thank Eric Ford for advice on implementation of the formula in Julia, and we thank Brett Morris and Ethan Kruse for advice on implementation of the formula in Python (requested by the referee).


\centerline{APPENDIX A}
\section*{A new approach to TTVs}\label{ttv_calculation}

Here we give the detailed derivation of the first-order solution presented in Section \ref{ttv_result}.

\subsection{TTVs from angular and radial variations}

We start with the equations of motion in Murray \& Dermott
(MD), 6.10-6.11, which are in heliocentric coordinates.  We use the 
index $i$ to label the planets, and
denote the inner planet with $i=1$ and outer with $i=2$.  
TTVs can be computed from the true longitudes, $\theta_1$ and $\theta_2$,
with respect to the star;  hence heliocentric coordinates are ideal
for transit timing computation.
We convert the equations of motion to polar coordinates, and isolate the
radial and longitudinal equations:
\begin{eqnarray}\label{forceequations1}
r_i^2 \ddot \theta_i + 2r_i\dot r_i\dot \theta_i &=& \frac{\partial}{\partial \theta_i} (U_i + {\cal R}_i) = \frac{\partial{\cal R}_i}{\partial \theta_i}=\dot l_i,\cr
\ddot{r_i} - r_i \dot \theta_i^2 &=& \frac{\partial}{\partial r_i} \left(U_i + {\cal R}_i\right).
\end{eqnarray}
The first equation can be rewritten as the time derivative of specific angular momentum, $l_i$,
and without the disturbing force it expresses conservation of angular momentum,
while the second equation includes centripetal acceleration in the radial direction
as the second term on the left hand side.
The term $U_i=G(m_\star+m_i)/r_i$ is the standard Keplerian potential,
while ${\cal R}_i$ is the disturbing function which reflects the
gravitational potential energy of planet-planet interactions.  Because
the planet-planet interactions lead to only small perturbations of
the base Keplerian orbits, we seek a solution to equations (\ref{forceequations1})
of the form: $r_i = r_{i,K}+\delta r_i$
and $\theta_i = \theta_{i,K}+\delta \theta_i$, where
$(r_{i,K},\theta_{i,K})$ is the unperturbed Keplerian polar coordinates
of the orbits, and $(\delta r_i,\delta \theta_i)$ are the
small perturbations.  We can then plug these solutions into the equations
of motion (\ref{forceequations1}) and expand in powers of $\delta r_i,
\delta \theta_i$, mass ratio, and eccentricity.

We normalize $r_i$ by $a_i$, which is the semi-major axis of the
unperturbed Keplerian orbit (we do not perturb $a_i$, $\varpi_i$, 
or $e_i$, so they are fixed at the mean, unperturbed values).  
Then $r_i/a_i = 1+\epsilon_i$, where $\epsilon_i$ is a 
dimensionless radial coordinate, so that $\dot r_i/a_i = \dot \epsilon_i$ and 
$\delta r_i/a_i = \delta \epsilon_i$.
The solution to the unperturbed orbits to first order in eccentricity is
$\theta_{i,K} = \lambda_i - Re\left[2\mathbbm{i}z_i^* e^{\mathbbm{i}\lambda_i}\right] = \lambda_i + 2e_i \sin{(\lambda_i-\varpi_i)}$
and $\epsilon_{i,K} = -Re\left[z_i^* e^{\mathbbm{i}\lambda_i}\right]$ where $\mathbbm{i}=\sqrt{-1}$,
$z_i=e_{i}e^{\mathbbm{i}\varpi_{i}}=e_i(\cos{\varpi_i}+\mathbbm{i}\sin{\varpi_i})$ 
is the complex eccentricity vector (LXW12), $z^*$ is the complex conjugate of $z$,
and $Re[.]$ is the real part. 

\subsection{Perturbed equations of motion}

The perturbed equations become:
\begin{eqnarray}
r_{i,K}^2\delta\ddot\theta_i  &=& -2r_{i,K}\ddot\theta_{i,K}\delta r_i - 2\dot r_{i,K}\dot\theta_{i,K}\delta r_i - 2r_{i,K}\dot\theta_{i,K}\delta\dot r_i - 2r_{i,K}\dot r_{i,K}\delta\dot\theta_i +\frac{\partial {\cal R}_i}{\partial \theta_i} ,\cr
\delta\ddot{r_i} &=& 2r_{i,K}\dot\theta_{i,K}\delta\dot\theta_i + \dot\theta_{i,K}^2\delta r_i + \frac{2G(m_\star+m_i)}{r_{i,K}^3}\delta r_i +\frac{\partial{\cal R}_i}{\partial r_i}.
\end{eqnarray}
In these equations, we have cancelled the terms for the unperturbed
Keplerian on both sides of the equation.
From hereon we will drop the `$K$' subscript from the unperturbed Keplerian
orbital elements.
To first order in eccentricity, the TTVs of the $i$th planet are:
\begin{equation}
\delta t_i = -n_i^{-1} (\delta \theta_i + 2\epsilon_i \delta \theta_i^{(0)}),
\end{equation}
where $\delta\theta_i^{(0)}$ is the perturbed solution to zeroth order in eccentricity.
To lowest order in eccentricity, $\epsilon_i = -e_i\cos{(\lambda_i-\varpi_i)}$, so:
\begin{eqnarray}
\epsilon_i &=& -\tfrac{1}{2}\left(e_ie^{\mathbbm{i}(\lambda_i-\varpi_i)}+e_ie^{-\mathbbm{i}(\lambda_i-\varpi_i)}\right)\cr
\dot \epsilon_i &=& -\tfrac{\mathbbm{i}n_i}{2}\left(e_ie^{\mathbbm{i}(\lambda_i-\varpi_i)}-e_ie^{-\mathbbm{i}(\lambda_i-\varpi_i)}\right).
\end{eqnarray}

\subsection{First order in eccentricity equations}

Let the specific angular
momentum equal $l_i = r_i^2 \dot \theta_i$. Since $l_i$ is a constant,
$\ddot \theta_i = -2l_i\dot r_i/r_i^3$. To first order in eccentricity,
$l_i = n_i a_i^2$, where $n_i = 2\pi/P_i$.  Substituting these into
the above equations, and expanding to first order in eccentricity, we
find
\begin{eqnarray}
\delta \ddot \epsilon_i -2n_i(1-\epsilon_i)\delta\dot\theta_i -n_i^2(3-10\epsilon_i)\delta \epsilon_i &=&
\frac{1}{a_i^2}\frac{\partial{\cal R}_i}{\partial \epsilon_i}\cr
(1+2\epsilon_i)\delta\ddot\theta_i - 2n_i\dot\epsilon_i\delta \epsilon_i+2n_i(1-\epsilon_i)\delta\dot \epsilon_i + 2\dot\epsilon_i\delta\dot\theta_i &=& \frac{1}{a_i^2}\frac{\partial{\cal R}_i}{\partial \theta_i}.
\end{eqnarray}
Note that the terms with $\epsilon_i$ or $\dot\epsilon_i$ are first order
in eccentricity; hence, the other quantities in these terms need to
only be expanded to zeroth order in eccentricity.  Denoting the zeroth-order
solutions as $\delta \epsilon_i^{(0)}$ and $\delta \theta_i^{(0)}$, we find the differential
equations governing the transit timing solution:
\begin{eqnarray}\label{inner_planet}
\delta\ddot \theta_i + 2 n_i\delta\dot \epsilon_i &=& -2 \left(\epsilon_i \delta \ddot \theta_i^{(0)} +\dot\epsilon_i\delta\dot\theta_i^{(0)}\right) + 2n_i\left(\dot\epsilon_i\delta \epsilon_i^{(0)} + \epsilon_i\delta\dot \epsilon_i^{(0)}\right)+  \frac{1}{a_i^2}\frac{\partial{\cal R}_i}{\partial \theta_i},\cr
\delta\ddot \epsilon_i - 3n_i^2\delta \epsilon_i - 2n_i\delta \dot \theta_i &=& -2n_i\epsilon_i\left(\delta\dot\theta_i^{(0)} + 5 n_i \delta \epsilon_i^{(0)}\right)+ \frac{1}{a_i^2}\frac{\partial{\cal R}_i}{\partial \epsilon_i}.
\end{eqnarray}
We assume that both sides of these equations are complex, and the final solution is found from
taking their real parts.


The quantity ${\cal R}_i$ is the disturbing function, which for the inner
planet can be broken into
two pieces: ${\cal R}_1 = \frac{Gm_2}{a_2}({\cal R}_D + \alpha {\cal R}_E)$
(MD 6.44), where ${\cal R_D} = a_2/\vert{\bf r}_2 - {\bf r}_1\vert$ and
${\cal R}_E = -(r_1/a_1)(a_2/r_2)^2 \cos{(\theta_1-\theta_2)}$.  
For the outer planet, there are also
two pieces, ${\cal R}_2 = \frac{Gm_1}{a_2}({\cal R}_D + \alpha^{-2} {\cal R}_I)$, where
${\cal R}_I = -(1+\epsilon_2)(1+\epsilon_1)^{-2} \cos{(\theta_1-\theta_2)}$ (MD 6.45).
As usual,
$\alpha = a_1/a_2 \approx (P_1/P_2)^{2/3}$.

\subsection{Expansion of the disturbing functions}

The expansion for ${\cal R}_D$ is given in MD 6.66.  We are considering
the plane-parallel case, so $\Psi =\cos{\psi}-\cos{(\theta_1-\theta_2)}= 0$, 
in which case we only need to include the $\propto\Psi^0$ term.  Also, we would 
like a solution that is first order in
eccentricity (of the unperturbed Keplerian orbit), so the term in brackets
in MD 6.66 needs to be expanded to second order in $\epsilon_{i}=\frac{r_{i}}{a_{i}}-1$
(noting again that $\epsilon_i$ is first order in eccentricity):
\begin{equation}
\sum_{l=0}^2 \frac{1}{l!} \sum_{k=0}^l {l \choose k} \epsilon_1^k \epsilon_2^{l-k}
A_{0,j,k,l-k} = A_{0,j,0,0}+ A_{0,j,1,0} \epsilon_1 + A_{0,j,0,1} \epsilon_2 + \frac{1}{2}A_{0,j,0,2}\epsilon_2^2 + A_{0,j,1,1}\epsilon_{1}\epsilon_2 + \frac{1}{2} A_{0,j,2,0}\epsilon_1^2,
\end{equation}
where $A_{0,j,m,n}=a_1^ma_2^n\frac{\partial^{m+n}}{\partial a_1^m \partial a_2^n}\left(a_2^{-1}b_{1/2}^{(j)}(\alpha)\right)$
(MD 6.63) and $b_{s}^{(j)}$ is a Laplace coefficient (MD 6.67).  We 
define $\tilde{A}_{jmn} = a_2 A_{0,j,m,n}$ to simplify the expressions below.

We rewrite ${\cal R}_D$ in complex notation (in the plane-parallel limit, expanded
to second order in $\epsilon_{i}$), giving:
\begin{eqnarray}
{\cal R}_D &=& {\cal R}_{D,0} + Re\Bigg[\sum_{j \ge 1} \left( \tilde{A}_{j00}+ \tilde{A}_{j10} \epsilon_1 + \tilde{A}_{j01} \epsilon_2 + \frac{1}{2}\tilde{A}_{j02}\epsilon_2^2 + \tilde{A}_{j11}\epsilon_{1}\epsilon_2 + \frac{1}{2} \tilde{A}_{j20}\epsilon_1^2\right) e^{\mathbbm{i}j(\theta_1-\theta_2)}\Bigg],\cr
{\cal R}_{D,0} &=& \frac{1}{2} \left(\tilde{A}_{000}+ \tilde{A}_{010} \epsilon_1 + \tilde{A}_{001} \epsilon_2 + \frac{1}{2}\tilde{A}_{002}\epsilon_2^2 + \tilde{A}_{011}\epsilon_{1}\epsilon_2 + \frac{1}{2} \tilde{A}_{020}\epsilon_1^2\right).
\end{eqnarray}
where 
we have used the fact that
$A_{0,j,k,l}=A_{0,-j,k,l}$ since $b_{s}^{(j)}=b_{s}^{(-j)}$. Likewise,
\begin{equation}
{\cal R}_E = -Re\bigg[(1+\epsilon_1)(1+\epsilon_2)^{-2} e^{\mathbbm{i}(\theta_1-\theta_2)}\bigg].
\end{equation}

Taking the derivative of ${\cal R}_D$ and ${\cal R}_E$
with respect to $\theta_1$ gives to first order in $\epsilon_1$:
\begin{equation}
\frac{1}{a_1^2}\frac{\partial{\cal R}_1}{\partial \theta_1} = \frac{1}{a_1^2}\delta \dot{l_1} = n_1^2 \mu_2 \alpha Re\left[
\sum_{j\ge 1} \left(\tilde{A}_{j00} + \tilde{A}_{j10} \epsilon_1 + \tilde{A}_{j01} \epsilon_2-\alpha (1+\epsilon_1-2\epsilon_2)\delta_{j1}\right)
\mathbbm{i}j e^{\mathbbm{i}j(\theta_1-\theta_2)} \right],
\end{equation}
where we have used $n_i^2=Gm_\star/a_i^3 = $ and $\mu_i=m_i/m_\star$.
The derivative of ${\cal R}_1$ with respect to $\epsilon_1$ is:
\begin{eqnarray}
\frac{1}{a_1^2}\frac{\partial {\cal R}_1}{\partial \epsilon_1}
&=& n_1^2 \mu_2 \alpha Re\Bigg[\frac{1}{2} \left(\tilde{A}_{010}+\tilde{A}_{011}\epsilon_2+\tilde{A}_{020}\epsilon_1\right)\cr
&+&\sum_{j\ge 1} \left(\tilde{A}_{j10}+\tilde{A}_{j11}\epsilon_2+\tilde{A}_{j20}\epsilon_1-\alpha(1-2\epsilon_2)\delta_{j1}\right)e^{\mathbbm{i}j(\theta_1-\theta_2)}  \Bigg].
\end{eqnarray}

For the outer planet,
\begin{equation}
\frac{1}{a_2^2}\frac{\partial{\cal R}_2}{\partial \theta_2} = \frac{1}{a_2^2}\delta \dot{l_2} = -n_2^2 \mu_1  Re\left[
\sum_{j\ge 1} \left(\tilde{A}_{j00} + \tilde{A}_{j10} \epsilon_1 + \tilde{A}_{j01} \epsilon_2-\alpha^{-2} (1+\epsilon_2-2\epsilon_1)\delta_{j1}\right)
\mathbbm{i}j e^{\mathbbm{i}j(\theta_1-\theta_2)} \right].
\end{equation}
The derivative of ${\cal R}_2$ with respect to $\epsilon_2$ is:
\begin{eqnarray}
\frac{1}{a_2^2}\frac{\partial {\cal R}_2}{\partial \epsilon_2}&=& n_2^2 \mu_1 Re\Bigg[
\frac{1}{2}\left(\tilde{A}_{001}+\tilde{A}_{011}\epsilon_1+\tilde{A}_{002}\epsilon_2\right) \cr
&+&\sum_{j\ge 1} \left(\tilde{A}_{j01}+\tilde{A}_{j11}\epsilon_1+\tilde{A}_{j02}\epsilon_2-\alpha^{-2}(1-2\epsilon_1)\delta_{j1}\right)e^{\mathbbm{i}j(\theta_1-\theta_2)}  \Bigg].
\end{eqnarray}

The angles and radii in the derivatives of the disturbing function can be expanded to first order in eccentricity, yielding:
\begin{eqnarray} \label{disturbing1}
\frac{1}{a_1^2}\frac{\partial {\cal R}_1}{\partial \theta_1} &=& n_1^2 \mu_2 \alpha Re\bigg[\sum_{j\ge 1} \mathbbm{i}j e^{\mathbbm{i}j\psi}\Big(\big(\tilde{A}_{j00}-\alpha\delta_{j1}\big)\cr
&+& \big(j\tilde{A}_{j00}-\tfrac{1}{2}\tilde{A}_{j10}-\tfrac{1}{2}\alpha\delta_{j1}\big)z_1^*e^{\mathbbm{i}\lambda_1}
+ \big(-j\tilde{A}_{j00}-\tfrac{1}{2}\tilde{A}_{j10}+\tfrac{3}{2}\alpha\delta_{j1}\big)z_1e^{-\mathbbm{i}\lambda_1}\cr
&+& \big(-j\tilde{A}_{j00}-\tfrac{1}{2}\tilde{A}_{j01}\big)z_2^*e^{\mathbbm{i}\lambda_2}
+ \big(j\tilde{A}_{j00}-\tfrac{1}{2}\tilde{A}_{j01}-2\alpha\delta_{j1}\big)z_2  e^{-\mathbbm{i}\lambda_2}\Big)\bigg].\cr
\frac{1}{a_1^2}\frac{\partial {\cal R}_1}{\partial \epsilon_1} &=& n_1^2 \mu_2 \alpha Re\Bigg[\tfrac{1}{2}\tilde{A}_{010}-\tfrac{1}{2}\tilde{A}_{020}
z_1e^{-\mathbbm{i}\lambda_1}
-\tfrac{1}{2}\tilde{A}_{011}z_2e^{-\mathbbm{i}\lambda_2} +\sum_{j\ge 1} e^{\mathbbm{i}j\psi}\Bigg(\big(\tilde{A}_{j10}-\alpha\delta_{j1}\big)\cr
&+& \left(j\tilde{A}_{j10}-\tfrac{1}{2}\tilde{A}_{j20}-\alpha\delta_{j1}\right)z_1^* e^{ i\lambda_1}
 +  \left(-j\tilde{A}_{j10}-\tfrac{1}{2}\tilde{A}_{j20}+\alpha\delta_{j1}\right)z_1   e^{-\mathbbm{i}\lambda_1}\cr
&+& \left(-j\tilde{A}_{j10}-\tfrac{1}{2}\tilde{A}_{j11}\right)z_2^* e^{ i\lambda_2}
 +  \left(j\tilde{A}_{j10}-\tfrac{1}{2}\tilde{A}_{j11}-2\alpha\delta_{j1}\right)z_2   e^{-\mathbbm{i}\lambda_2}\Bigg)\Bigg].
\end{eqnarray}
where $\psi = \lambda_1-\lambda_2$. 
We have also combined the $j=0$ eccentricity terms since $Re(z_i^*e^{\mathbbm{i}\lambda_i}) = Re(z_ie^{-\mathbbm{i}\lambda_i})= \tfrac{1}{2}(z_i^*e^{\mathbbm{i}\lambda_i}+z_ie^{-\mathbbm{i}\lambda_i})$.

For the outer planet,
\begin{eqnarray} 
\frac{1}{a_2^2}\frac{\partial {\cal R}_2}{\partial \theta_2} &=& -n_2^2 \mu_1 Re\bigg[\sum_{j\ge 1} \mathbbm{i}j e^{\mathbbm{i}j\psi}\Big(\big(\tilde{A}_{j00}-\alpha^{-2}\delta_{j1}\big)\cr
&+& \big(j\tilde{A}_{j00}-\tfrac{1}{2}\tilde{A}_{j10}-2\alpha^{-2}\delta_{j1}\big)z_1^*e^{\mathbbm{i}\lambda_1}
+ \big(-j\tilde{A}_{j00}-\tfrac{1}{2}\tilde{A}_{j10}\big)z_1e^{-\mathbbm{i}\lambda_1}\cr
&+& \big(-j\tilde{A}_{j00}-\tfrac{1}{2}\tilde{A}_{j01}+\tfrac{3}{2}\alpha^{-2}\delta_{j1}\big)z_2^*e^{\mathbbm{i}\lambda_2}
+ \big(j\tilde{A}_{j00}-\tfrac{1}{2}\tilde{A}_{j01}-\tfrac{1}{2}\alpha^{-2}\delta_{j1}\big)z_2  e^{-\mathbbm{i}\lambda_2}\Big)\bigg].\cr
\frac{1}{a_2^2}\frac{\partial {\cal R}_2}{\partial \epsilon_2} &=& n_2^2 \mu_1 Re\Bigg[\tfrac{1}{2}\tilde{A}_{001} - \tfrac{1}{2}\tilde{A}_{011}z_1^* e^{\mathbbm{i}\lambda_1}
- \tfrac{1}{2}\tilde{A}_{002}z_2^* e^{\mathbbm{i}\lambda_2} +
\sum_{j\ge 1} e^{\mathbbm{i}j\psi}\Bigg(\big(\tilde{A}_{j01}-\alpha^{-2}\delta_{j1}\big)\cr
&+& \left(j\tilde{A}_{j01}-\tfrac{1}{2}\tilde{A}_{j11}-2\alpha^{-2}\delta_{j1}\right)z_1^* e^{ i\lambda_1}
 +  \left(-j\tilde{A}_{j01}-\tfrac{1}{2}\tilde{A}_{j11}\right)z_1   e^{-\mathbbm{i}\lambda_1}\cr
&+& \left(-j\tilde{A}_{j01}-\tfrac{1}{2}\tilde{A}_{j02}+\alpha^{-2}\delta_{j1}\right)z_2^* e^{ i\lambda_2}
 +  \left(j\tilde{A}_{j01}-\tfrac{1}{2}\tilde{A}_{j02}-\alpha^{-2}\delta_{j1}\right)z_2   e^{-\mathbbm{i}\lambda_2}\Bigg)\Bigg].
\end{eqnarray}

\subsection{Trial solution}

The derivatives of the complex disturbing function contain terms that are proportional
to $e^{\mathbbm{i}j\psi}$, $e^{\mathbbm{i}j\psi}e_{i}e^{\pm \mathbbm{i}(\lambda_{i}-\varpi_{i})}$
for $j\ge 0$. We treat these terms as harmonic driving terms, and solve the inhomogeneous
partial differential equations term by term.
We expand the complex solutions for $\delta \theta_1$ and $\delta \epsilon_1$ as trial
solutions:
\begin{eqnarray}
\delta \epsilon_1 &=& 
\delta\epsilon_{1,0}^{(+2)} e_2 e^{\mathbbm{i}(\lambda_2-\varpi_2)}
+ \sum_{j\ge 1} \delta\epsilon_{1,j} e^{\mathbbm{i}j\psi} \cr
\delta \theta_1 &=&  
\delta\theta_{1,0}^{(+2)} e_2 e^{\mathbbm{i}(\lambda_2-\varpi_2)}
+\sum_{j\ge 1} \delta\theta_{1,j} e^{\mathbbm{i}j\psi}\cr
\delta \epsilon_{1,j} &=& \delta \epsilon_{1,j}^{(0)} + \sum_{k=1,2} \left(\delta \epsilon_{1,j}^{(+k)} e_ke^{\mathbbm{i}(\lambda_k-\varpi_k)} + \delta \epsilon_{1,j}^{(-k)}e_ke^{-\mathbbm{i}(\lambda_k-\varpi_k)}\right),\cr
\delta \theta_{1,j} &=& \delta \theta_{1,j}^{(0)} + \sum_{k=1,2} \left(\delta\theta_{1,j}^{(+k)} e_ke^{\mathbbm{i}(\lambda_k-\varpi_k)} + \delta \theta_{1,j}^{(-k)}e_ke^{-\mathbbm{i}(\lambda_k-\varpi_k)}\right),
\end{eqnarray}
and
\begin{eqnarray}
\delta \epsilon_2 &=& 
\delta\epsilon_{2,0}^{(+1)} e_1 e^{\mathbbm{i}(\lambda_1-\varpi_1)}
+ \sum_{j\ge 1} \delta\epsilon_{2,j} e^{\mathbbm{i}j\psi} \cr
\delta \theta_2 &=&  
\delta\theta_{1,0}^{(+1)} e_1 e^{\mathbbm{i}(\lambda_1-\varpi_1)}
+\sum_{j\ge 1} \delta\theta_{2,j} e^{\mathbbm{i}j\psi}\cr
\delta \epsilon_{2,j} &=& \delta \epsilon_{2,j}^{(0)} + \sum_{k=1,2} \left(\delta \epsilon_{2,j}^{(+k)} e_ke^{\mathbbm{i}(\lambda_k-\varpi_k)} + \delta \epsilon_{2,j}^{(-k)}e_ke^{-\mathbbm{i}(\lambda_k-\varpi_k)}\right),\cr
\delta \theta_{2,j} &=& \delta \theta_{2,j}^{(0)} + \sum_{k=1,2} \left(\delta\theta_{2,j}^{(+k)} e_ke^{\mathbbm{i}(\lambda_k-\varpi_k)} + \delta \theta_{2,j}^{(-k)}e_ke^{-\mathbbm{i}(\lambda_k-\varpi_k)}\right).
\end{eqnarray}
We also define the solutions to zeroth order in eccentricity as:
\begin{eqnarray}
\delta \epsilon_i^{(0)} &=& \sum_{j\ge 1} \epsilon_{i,j}^{(0)} e^{\mathbbm{i}j\psi} \cr
\delta \theta_i^{(0)} &=& \sum_{j\ge 1} \theta_{i,j}^{(0)} e^{\mathbbm{i}j\psi}.
\end{eqnarray}
Then, the (real) transit timing variations are equal to
\begin{eqnarray}
\delta t_1 &=& -n_1^{-1} Re(\delta \theta_1 + 2 \epsilon_1 \delta \theta_1^{(0)})\cr
 &=& \delta t_{1,0}^{(-2)} + \sum_{j\ge 1} \left[\delta t_{1,j}^{(0)} + \sum_{k=1,2} \left(\delta t_{1,j}^{(+k)} + \delta t_{1,j}^{(-k)}\right) \right],\cr
\delta t_{1,j}^{(0)} &=& -n_1^{-1} Re(\delta \theta_{1,j}^{(0)} e^{\mathbbm{i}j\psi})\cr
\delta t_{1,j}^{(\pm k)} &=& -n_1^{-1} Re\left(\left(\delta \theta_{1,j}^{(\pm k)}-\delta \theta_{1,j}^{(0)} \delta_{k1}\right) e_k e^{\mathbbm{i}(j\psi\pm (\lambda_k-\varpi_k))}\right),
\end{eqnarray}
and
\begin{eqnarray}
\delta t_2 &=& -n_2^{-1} Re(\delta \theta_2 + 2 \epsilon_2 \delta \theta_2^{(0)})\cr
 &=& \delta t_{2,0}^{(+1)} + \sum_{j\ge 1} \left[\delta t_{2,j}^{(0)} + \sum_{k=1,2} \left(\delta t_{2,j}^{(+k)} + \delta t_{2,j}^{(-k)}\right) \right],\cr
\delta t_{2,j}^{(0)} &=& -n_2^{-1} Re(\delta \theta_{2,j}^{(0)} e^{\mathbbm{i}j\psi})\cr
\delta t_{2,j}^{(\pm k)} &=& -n_2^{-1} Re\left(\left(\delta \theta_{2,j}^{(\pm k)}-\delta \theta_{2,j}^{(0)} \delta_{k2}\right) e_k e^{\mathbbm{i}(j\psi\pm (\lambda_k-\varpi_k))}\right),
\end{eqnarray}

\subsection{Inner planet coefficients}

Substituting the $(j,{0,\pm k})$ trial solutions into the above differential equations, to zeroth order
in eccentricity we find for the inner planet:
\begin{equation}\label{first_matrix}
\begin{pmatrix}
-\beta_j^2 &  2\mathbbm{i}\beta_j \\
-2\mathbbm{i}\beta_j & -(\beta_j^2+3)
\end{pmatrix}
\begin{pmatrix}
\delta \theta_{1,j}^{(0)}\\
\delta \epsilon_{1,j}^{(0)}
\end{pmatrix}
= \mu_2 \alpha
\begin{pmatrix}
\mathbbm{i}j(\tilde{A}_{j00}-\alpha \delta_{j1})\\
\tilde{A}_{j10} - \alpha \delta_{j1}
\end{pmatrix}.
\end{equation}
where $\beta_j = j (n_1-n_2)/n_1 = j(1-\alpha^{3/2})$ and we have divided the equations
by $n_1^2$ to make dimensionless.

Similarly we can write down the equations for the coefficients
to first order in $e_1$.  Note that to solve for $\delta t_{1,j}^{(\pm 1)}$, we need to
compute $\delta \theta_{1,j}^{(\pm 1)} - \delta \theta_{1,j}^{(0)}$.  Hence
we can subtract the matrix on the left times the vector
$\left\{\delta \theta_{1,j}^{(0)},0\right\}$ from both sides of the equation.  This may
be rewritten in dimensionless form as:
\begin{eqnarray}\label{inner_e1}
&&
\begin{pmatrix}
-(\beta_j\pm 1)^2 &  2\mathbbm{i}(\beta_j\pm 1) \\
-2\mathbbm{i}(\beta_j \pm 1)& -((\beta_j\pm 1)^2+3)
\end{pmatrix}
\begin{pmatrix}
\delta \theta_{1,j}^{(\pm 1)} - \delta \theta_{1,j}^{(0)}\\
\delta \epsilon_{1,j}^{(\pm 1)}
\end{pmatrix}\cr
&=&
\begin{pmatrix}
1\pm\beta_j  & -\mathbbm{i}(\beta_j\pm 1)\\
\mathbbm{i}(3\beta_j\pm 2) & 5
\end{pmatrix}
\begin{pmatrix}
\delta \theta_{1,j}^{(0)}\\
\delta \epsilon_{1,j}^{(0)}
\end{pmatrix}\cr
&+&
\mu_2 \alpha
\begin{pmatrix}
\mathbbm{i}j\left(\pm j\tilde{A}_{j00}-\tfrac{1}{2}\tilde{A}_{j10}+\tfrac{1}{2}\alpha \delta_{j1}(1\mp 2)\right)\\
\pm j \tilde{A}_{j10} -\tfrac{1}{2}\tilde{A}_{j20}\mp \alpha \delta_{j1} 
\end{pmatrix}.
\end{eqnarray}
The solutions of equation \ref{first_matrix} for $(\delta\theta_{1,j}^{(0)},\delta\epsilon_{1,j}^{(0)})$
may be plugged into the first term on the right hand side of this equation to solve for the
first order eccentricity terms.

To first order in $e_2$ in dimensionless form (for $j \ge 1$),
\begin{equation}
\begin{pmatrix}
-\eta_\pm^2 &  2\mathbbm{i}\eta_\pm \\
-2\mathbbm{i}\eta_\pm & -(\eta_\pm^2+3)
\end{pmatrix}
\begin{pmatrix}
\delta \theta_{1,j}^{(\pm 2)}\\
\delta \epsilon_{1,j}^{(\pm 2)}
\end{pmatrix}
=
\mu_2 \alpha
\begin{pmatrix}
\mathbbm{i}j\left(\mp j\tilde{A}_{j00}-\tfrac{1}{2}\tilde{A}_{j01}-(1\mp 1)\alpha \delta_{j1}\right)\\  
\mp j \tilde{A}_{j10} -\tfrac{1}{2}\tilde{A}_{j11}- (1\mp 1) \alpha \delta_{j1}   
\end{pmatrix},
\end{equation}
where $\eta_{\pm} = \beta_j \pm \alpha^{3/2} = j(n_1-n_2)/n_1\pm n_2/n_1 = j(1-\alpha^{3/2})\pm \alpha^{3/2}$
and $\tilde{A}_{j11} = -(2\tilde{A}_{j10} +\tilde{A}_{j20})
= -2 \alpha \partial b_{1/2}^{(j)}/\partial \alpha -\alpha^2 \partial^2 b_{1/2}^{(j)}/\partial \alpha^2$.
Note that for $j=1$ the $\eta_+$ term becomes $\eta_+ = 1$.  The frequency dependence of this term is
at the Keplerian frequency of the inner planet, $n_1$, and the determinant of the left hand matrix
becomes zero as the second row of the matrix is equal to $2\mathbbm{i}$ times the first row.  This singularity
occurs due to the fact that the equations become those of a resonantly driven oscillator, which means
that the amplitude grows linearly with time (or, equivalently on short timescales, the Keplerian frequency
is shifted).  This term is not relevant for transit-timing analyses as it occurs at the frequency of
the transiting planet and grows on the secular timescale;  consequently we will drop this term for
now and discuss below in appendix B.

For $j=0$,
\begin{equation}
\begin{pmatrix}
-\alpha^3 &   -2\mathbbm{i}\alpha^{3/2} \\
2\mathbbm{i}\alpha^{3/2} & -(\alpha^3+3)
\end{pmatrix}
\begin{pmatrix}
\delta \theta_{1,0}^{(-2)}\\
\delta \epsilon_{1,0}^{(-2)}
\end{pmatrix}
=
\mu_2 \alpha
\begin{pmatrix}
0\\
-\tfrac{1}{2} \tilde{A}_{011}
\end{pmatrix}.
\end{equation}

\subsection{Outer planet coefficients}

Defining $\kappa_j = j(n_1-n_2)/n_2 = \alpha^{-3/2}\beta_j$, dividing
this equation by $n_2^2$ gives the dimensionless form of:
\begin{equation}
\begin{pmatrix}
-\kappa_j^2 & 2\mathbbm{i}\kappa_j \\
-2\mathbbm{i}\kappa_j & -(\kappa_j^2+3)
\end{pmatrix}
\begin{pmatrix}
\delta \theta_{2,j}^{(0)}\\
\delta \epsilon_{2,j}^{(0)}
\end{pmatrix}
=
\mu_1
\begin{pmatrix}
-\mathbbm{i}j(\tilde{A}_{j00}-\alpha^{-2}\delta_{j1})\\
\tilde{A}_{j01}-\alpha^{-2}\delta_{j1}
\end{pmatrix},
\end{equation}
and for the equations to first order in $e_1$ for $j \ge 1$,

\begin{equation}\label{outer_e2}
\begin{pmatrix}
-\xi_\pm^2 & 2\mathbbm{i}\xi_\pm \\
-2\mathbbm{i}\xi_\pm & -(\xi_\pm^2+3)
\end{pmatrix}
\begin{pmatrix}
\delta \theta_{2,j}^{(\pm 1)}\\
\delta \epsilon_{2,j}^{(\pm 1)}
\end{pmatrix}
=
\mu_1
\begin{pmatrix}
-\mathbbm{i}j(\pm j\tilde{A}_{j00}-\tfrac{1}{2}\tilde{A}_{j10}-(1\pm 1)\alpha^{-2}\delta_{j1})\\
\pm j \tilde{A}_{j01}-\tfrac{1}{2}\tilde{A}_{j11}-(1\pm 1)\alpha^{-2}\delta_{j1}
\end{pmatrix},
\end{equation}
where $\xi_\pm = \kappa_j \pm \alpha^{-3/2}$,
while for $j=0$,
\begin{equation}
\begin{pmatrix}
-\alpha^{-3} &  2\mathbbm{i}\alpha^{-3/2} \\
-2\mathbbm{i}\alpha^{-3/2} & -(\alpha^{-3}+3)
\end{pmatrix}
\begin{pmatrix}
\delta \theta_{2,0}^{(+1)}\\
\delta \epsilon_{2,0}^{(+1)}
\end{pmatrix}
=
\mu_1
\begin{pmatrix}
0\\
-\tfrac{1}{2} \tilde{A}_{011}
\end{pmatrix},
\end{equation}
and for first order in $e_2$,
\begin{eqnarray}\label{last_matrix}
&&\begin{pmatrix}
-(\kappa_j\pm 1)^2 & 2\mathbbm{i}(\kappa_j\pm 1) \\
-2\mathbbm{i}(\kappa_j \pm 1) & -((\kappa_j\pm 1)^2+3)
\end{pmatrix}
\begin{pmatrix}
\delta \theta_{2,j}^{(\pm 2)}-\delta \theta_{2,j}^{(0)}\\
\delta \epsilon_{2,j}^{(\pm 2)}
\end{pmatrix}\cr
&=&
\begin{pmatrix}
1\pm \kappa_j & -\mathbbm{i}(\kappa_j\pm 1)\\
\mathbbm{i}(3\kappa_j \pm 2) & 5
\end{pmatrix}
\begin{pmatrix}
\delta \theta_{2,j}^{(0)}\\
\delta \epsilon_{2,j}^{(0)}\\
\end{pmatrix}
+\mu_1
\begin{pmatrix}
-\mathbbm{i}j(\mp j\tilde{A}_{j00}-\tfrac{1}{2}\tilde{A}_{j01}+\tfrac{1}{2}(1\pm 2)\alpha^{-2}\delta_{j1})\\
\mp j \tilde{A}_{j01}-\tfrac{1}{2}\tilde{A}_{j02}\pm \alpha^{-2}\delta_{j1}
\end{pmatrix}.
\end{eqnarray}

The function $u$ originates from the inversion of the matrices
in the left hand side of equations (\ref{first_matrix})-(\ref{last_matrix}), 
which each have the form of:
\begin{equation}
\begin{pmatrix}
-\gamma^2 & 2\mathbbm{i}\gamma \\
-2\mathbbm{i}\gamma & -(\gamma^2+3),
\end{pmatrix}
\end{equation}
where $\gamma$ is the dimensionless frequency in units of the
orbital frequency of the transiting planet.  
Since TTVs only depend
upon $\delta \theta$, then only the first term in the inverse of
this matrix times the right hand side of the equation gives the
function $u$.   These terms are driven by the disturbing function
only.

The function $v_\pm$ results from the driving terms caused by 
appearance of the zeroth-order eccentricity equation on the
right hand side of the linearized equations (\ref{inner_planet}).  
These can be expressed as the inverse of
the zeroth-order matrix times the coefficients of the disturbing
function driving the TTVs at zeroth order;  the first term on
the right hand sides of equations (\ref{inner_e1}) and (\ref{outer_e2})
times the inverse of the matrix on the left hand side yield the
functions $v_\pm$.

The expressions for $u$ and $v_\pm$ result from the coefficients in 
equations (\ref{first_matrix}-\ref{last_matrix}) which can be found 
by inverting the matrices.  Note that the coefficients of
$\delta\theta_i$ are imaginary, while $\delta \epsilon_i$ are real;  thus,
$\delta\theta_i$, and hence $\delta t_i$, will always have a sine dependence,
while $\delta\epsilon_i$ will always have a cosine dependence.

As with the inner planet, for $j=1$ the $\xi_-$ term becomes $\xi_- = 1$.  The frequency
dependence of this term is at the Keplerian frequency of the outer planet, $n_2$, and
hence the determinant of the left hand matrix
becomes zero as the second row of the matrix is equal to $2\mathbbm{i}$ times the first row (as
occurs for the inner planet).  We will drop this term for
now and discuss next in appendix B.


\centerline{APPENDIX B}
\section*{Secular terms}\label{secular}

In the foregoing analysis we neglected the presence of secular terms, which
in the disturbing function appear at zero frequency as well as at the Keplerian
frequency of the planet that is being perturbed.  These terms cause corrections
of ${\cal O}(\mu^1)$ to the ephemeris of the planet, and thus cause an error
of ${\cal O}(\mu^1)$ to the computation of $\alpha$ from the best-fit mean period.
The correction to $\alpha$ affects the coefficients of the TTVs at
order ${\cal O}(\mu^2)$, and so it can be neglected for the purposes of the
first order in eccentricity transit
timing solution.  However, the solution we
present here may have other applications, such as for radial-velocity planets
or astrometric motion, which are not aliased at the orbital frequency of
the planets, and so these secular terms enter at the ${\cal O}(\mu^1)$ level. 
In this appendix we compute these secular terms to first order in $\mu$
and $e$.

In the equations of motion for the inner planet, to include the secular and
Keplerian frequency terms, we will use angular momentum, $l_1$ in lieu of
angle $\theta_1$.  The equations of motion become:
\begin{eqnarray}
\dot l_1 &=&  \frac{\partial{\cal R}_1}{\partial \theta_1},\cr
\ddot \epsilon_1 - \frac{l_1^2}{a_1^4(1+\epsilon_1)^3} &=& -n_1^2(1+\epsilon_1)^{-2} + \frac{1}{a_1^2}  \frac{\partial{\cal R}_1}{\partial \epsilon_1},
\end{eqnarray}
where we have made the substitution $r_1=a_1(1+\epsilon_1)$ into equations (\ref{forceequations1}) and we have
divided by $a_1$.  In equation \ref{disturbing1} we keep only the secular terms and terms at frequency $n_1$,
giving:
\begin{eqnarray}
\dot l_1 &=&  -n_1^2 \mu_2 a_1^2 \alpha Re\big[\mathbbm{i}(\tilde{A}_{100}+\tfrac{1}{2} \tilde{A}_{101}) z_2^* e^{\mathbbm{i}\lambda_1} \big]\cr
\ddot \epsilon_1 - \frac{l_1^2}{a_1^4(1+\epsilon_1)^3} &=& -\frac{n_1^2}{(1+\epsilon_1)^{2}}
+ n_1^2 \mu_2 \alpha Re\Big[\tfrac{1}{2}\tilde{A}_{010} - \tfrac{1}{2} \tilde{A}_{020}z_1^* e^{\mathbbm{i}\lambda_1}
-(\tilde{A}_{110}+\tfrac{1}{2}\tilde{A}_{111})z_2^* e^{\mathbbm{i}\lambda_1}\Big].
\end{eqnarray}
for the inner planet, and similarly for the outer planet
\begin{eqnarray}
\dot l_2 &=&  -n_2^2 \mu_1 a_2^2 Re\left[\mathbbm{i}(\tilde{A}_{100}+\tfrac{1}{2} \tilde{A}_{110}) z_1^* e^{\mathbbm{i}\lambda_2} \right]\cr
\ddot \epsilon_2 - \frac{l_2^2}{a_2^4(1+\epsilon_2)^3} &=& -\frac{n_2^2}{(1+\epsilon_2)^{2}}
+ \tfrac{1}{2}n_2^2 \mu_1 Re\left[\tilde{A}_{001} - \tilde{A}_{002}z_2^* e^{\mathbbm{i}\lambda_2}
-(2\tilde{A}_{101}+\tilde{A}_{111})z_1^* e^{\mathbbm{i}\lambda_2}\right],
\end{eqnarray}
where we have taken the complex conjugate since this does not change the real component.

The solution for the inner planet's angular momentum to first order in eccentricity is:
\begin{equation}
l_1 = n_1 a_1^2 \Big(1 -\tfrac{1}{4} \mu_2 \alpha \tilde{A}_{010} - \mu_2 \alpha (\tilde{A}_{100}+\tfrac{1}{2}\tilde{A}_{101})Re\big[z_2^* e^{\mathbbm{i}\lambda_1} \big]\Big).
\end{equation}
This can be substituted into the equation for
$\epsilon_1$, keeping terms of order eccentricity, to obtain:
\begin{eqnarray}
\ddot \epsilon_1 + n_1^2 \epsilon_1 &=& -n_1^2 \mu_2 \alpha Re[g_1 e^{\mathbbm{i}\lambda_1}]\cr
g_1 &=& \tfrac{1}{2}(3\tilde{A}_{010}+\tilde{A}_{020})z_1^*+(2\tilde{A}_{100}+\tilde{A}_{101}+\tilde{A}_{110}
+\tfrac{1}{2}\tilde{A}_{111})z_2^*.
\end{eqnarray}
Note that we have chosen the constant of integration in $l_1$ to cancel the constant term in
the disturbing function derivative that appears in the equation for $\epsilon_1$ so that
the $\epsilon_1$ does not have an offset.  This is because we prefer to specify the value
of the semi-major axis in the initial conditions.

The solution to this equation is: 
\begin{equation}
\epsilon_{1,sec} = Re\Big[\left(-z_1^* + \mathbbm{i}\mu_2 \alpha g_1 \lambda_1/2\right)e^{\mathbbm{i}\lambda_1}\Big].
\end{equation}
Note that this solution grows in amplitude linearly with time;  however, the growth is slow,
occuring on the secular timescale times the inverse of the eccentricity.

With these solutions in hand, we can solve for $\dot \theta_1 = l_1/r_1^2 \approx l_1 a_1^{-2}(1-2\epsilon_1)+\mathcal{O}(e_1^2)$.
We then integrate this with respect to time, giving:
\begin{eqnarray} 
\theta_{1,sec} &=& \lambda_1(1 -\tfrac{1}{4}\mu_2 \alpha \tilde{A}_{010}) + Re\Big[-2\mathbbm{i} z_1^* e^{\mathbbm{i}\lambda_1}-\mu_2 \lambda_1 \alpha g_1 e^{\mathbbm{i}\lambda_1}\cr
&-& \mathbbm{i} \mu_2 \alpha \left\{g_1-(\tilde{A}_{100}+\tfrac{1}{2}\tilde{A}_{101})z_2^*-\tfrac{1}{2}\tilde{A}_{010}z_1^* \right\}e^{\mathbbm{i}\lambda_1}\Big].
\end{eqnarray}

A similar solution can be derived for the outer planet, with:
\begin{equation}
l_2 = n_2 a_2^2 \Big(1-\tfrac{1}{4}\mu_1 \tilde{A}_{001} - \mu_1 (\tilde{A}_{100}+\tfrac{1}{2}\tilde{A}_{110})Re\left[z_1^* e^{\mathbbm{i}\lambda_2}\right]\Big).
\end{equation}
As before, this can be substituted into the equation for $\epsilon_2$:
\begin{eqnarray}
\ddot \epsilon_2 + n_2^2 \epsilon_2 &=& -n_2^2 \mu_1 Re[g_2 e^{\mathbbm{i}\lambda_2}]\cr
g_2 &=& \tfrac{1}{2}(3\tilde{A}_{001}+\tilde{A}_{002})z_2^*+(2\tilde{A}_{100}+\tilde{A}_{110}+\tilde{A}_{101}
+\tfrac{1}{2}\tilde{A}_{111})z_1^*.
\end{eqnarray}
This equation has solution
\begin{equation}
\epsilon_{2,sec} = Re\Big[\left(-z_2^* +\mathbbm{i}g_2 \mu_1 \lambda_2/2\right)e^{\mathbbm{i}\lambda_2}\Big].
\end{equation}

Substituting this into the relation  $\dot \theta_2 = l_2/r_2^2 \approx l_2 a_2^{-2}(1-2\epsilon_2)+\mathcal{O}(e_2^2)$ and
integrating $\dot \theta_2$ with respect to time yields:
\begin{eqnarray}
\theta_{2,sec} &=& \lambda_2(1 -\tfrac{1}{4}\mu_1 \tilde{A}_{001}) + Re\Big[-2\mathbbm{i} z_2^* e^{\mathbbm{i}\lambda_2}-\mu_1 \lambda_2 g_2 e^{\mathbbm{i}\lambda_2}\cr
&-& \mathbbm{i} \mu_1 \left\{g_2-(\tilde{A}_{100}+\tfrac{1}{2}\tilde{A}_{110})z_1^*-\tfrac{1}{2}\tilde{A}_{020}z_2^* \right\}e^{\mathbbm{i}\lambda_2}\Big].
\end{eqnarray}
We have verified these solutions by plugging them back into the
differential equations, both analytically and numerically.
The solutions for $\theta_{i,sec}$ can be transformed to timing
variations with $\delta t_{i,sec}=-\dot \theta_{i,sec}^{-1}(\theta_{i,sec}-\lambda_i)$.

\end{document}